\documentclass[aip, amsmath, amssymb, reprint]{revtex4-1}

\usepackage{graphicx}
\usepackage{dcolumn}
\usepackage{bm}

\usepackage[utf8]{inputenc}
\usepackage[T1]{fontenc}
\usepackage{mathptmx}
\usepackage{etoolbox}

\usepackage{bbm}
\usepackage{physics}

\makeatletter
\def\@email#1#2{%
 \endgroup
 \patchcmd{\titleblock@produce}
  {\frontmatter@RRAPformat}
  {\frontmatter@RRAPformat{\produce@RRAP{*#1\href{mailto:#2}{#2}}}\frontmatter@RRAPformat}
  {}{}
}%
\makeatother

\makeatletter
\newsavebox{\@brx}
\newcommand{\llangle}[1][]{\savebox{\@brx}{\(\m@th{#1\langle}\)}%
  \mathopen{\copy\@brx\mkern2mu\kern-0.9\wd\@brx\usebox{\@brx}}}
\newcommand{\rrangle}[1][]{\savebox{\@brx}{\(\m@th{#1\rangle}\)}%
  \mathclose{\copy\@brx\mkern2mu\kern-0.9\wd\@brx\usebox{\@brx}}}
\makeatother

\newcommand{\exeter}{Department of Physics and Astronomy, University of Exeter, Stocker Road, Exeter EX4 4QL, United Kingdom.}
\newcommand{\potsdam}{Institut f\"{u}r Physik und Astronomie, University of Potsdam, 14476 Potsdam, Germany.}

\newcommand{\Gibbs}{\op{\tau}_\mathrm{Gibbs}}
\newcommand{\MF}{\op{\tau}_\mathrm{MF}}
\newcommand{\HHO}{\op{H_0}}
\newcommand{\sys}{\mathrm{S}}
\newcommand{\bath}{\mathrm{B}}
\newcommand{\sysbath}{\mathrm{SB}}
\newcommand{\inter}{\mathrm{I}}
\newcommand{\tot}{\mathrm{tot}}
\newcommand{\cl}{\mathrm{cl}}
\newcommand{\qu}{\mathrm{qu}}
\newcommand{\sto}{\mathrm{sto}}
\newcommand{\lap}[1]{\tilde{#1}}
\newcommand{\invlap}{\mathcal{L}^{-1}}
\newcommand{\fourier}[1]{\hat{#1}}
\renewcommand{\Re}{\mathrm{Re}}
\renewcommand{\Im}{\mathrm{Im}}

\renewcommand{\op}[1]{\pmb{#1}}
\newcommand{\kB}{k_\mathrm{B}}
\newcommand{\kBT}{k_\mathrm{B}T}
\newcommand{\wHO}{\Omega}
\newcommand{\wHOS}{\bar{\Omega}}
\newcommand{\wLor}{\omega_0}

\newcommand{\GammaLor}{\Gamma}
\newcommand{\spd}{J}
\newcommand{\noise}{N}
\newcommand{\noiseCl}{N^{\cl}}
\newcommand{\noiseQu}{N^{\qu}}
\newcommand{\memker}{K}
\newcommand{\psd}{P}

\newcommand{\noisei}{N_{T_\alpha}}
\newcommand{\noiseker}{\nu}
\newcommand{\lambdaUnits}{\Omega^2 m^{1/2}}
\newcommand{\vect}[1]{\underline{#1}}
\newcommand{\mat}[1]{\underline{\underline{#1}}}

\begin{document}

\title{Stochastic simulation of dissipative quantum oscillators}


\author{C. R. Hogg}
\email{c.r.hogg@exeter.ac.uk}
\affiliation{\exeter}

\author{J. Glatthard}
\affiliation{\exeter}

\author{F. Cerisola}
\affiliation{\exeter}

\author{J. Anders}
\affiliation{\exeter}
\affiliation{\potsdam}

\date{\today}
 
\begin{abstract}
Generic open quantum systems are notoriously difficult to simulate unless one looks at specific regimes. In contrast, classical dissipative systems can often be effectively described by stochastic processes, which are generally less computationally expensive. Here, we use the paradigmatic case of a dissipative quantum oscillator to give a pedagogic introduction into the modelling of open quantum systems using quasiclassical methods, i.e. classical stochastic methods that use a `quantum' noise spectrum to capture the influence of the environment on the system. Such quasiclassical methods have the potential to offer insights into the impact of the quantum nature of the environment on the dynamics of the system of interest whilst still being computationally tractable.
\end{abstract}

\maketitle

\section{Introduction}

Understanding how nanoscale systems interact and exchange energy with their environment is a crucial question for the development of quantum technologies. For example, characterizing the influence of the environment on the qubits that comprise quantum computers \cite{verstraete2009, ashhab2006}, or in quantum metrology, where one needs to accurately model the interaction of a probe with a sample, e.g. to measure its temperature \cite{depasquale2018, mehboudi2019}. The system of interest and its environment collectively form an `open system' \cite{breuer_book}. At the quantum level, the interaction between the system and environment cannot be ignored \cite{rivas2010}, and being able to model the full open system + environment dynamics is essential to understand the dynamics of the system of interest. As such, there has been a sustained effort to advance the framework for modelling dissipation in the quantum realm \cite{weiss_book}. One difficulty is that the system may be intricately correlated with the environment \cite{miller2018}, and obtaining the reduced state of the system alone tends to be a difficult task and computationally expensive \cite{strathearn2018, gribben2022, tanimura1989, jin2008, prior2010, chin2010, strasberg2016, nazir2018}. In contrast, one can effectively simulate \emph{classical} systems that are strongly coupled to their environment \cite{ford1965, zwanzig1973}.

\begin{figure}
    \centering
    \includegraphics{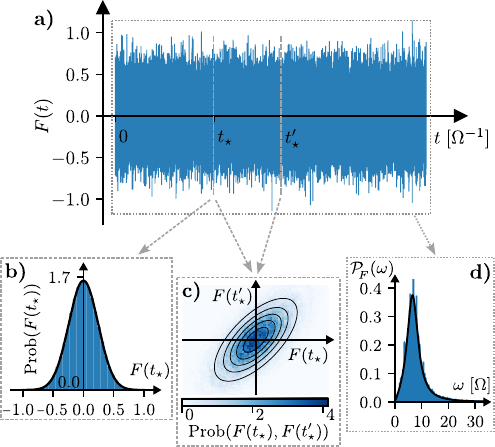}
    \caption{\textbf{Stochastic force}. Panel \textbf{a)} shows a single realization of the stochastic force as a function of time, $F(t)$, which has the Gaussian probability distribution given in \textbf{b)}. Panel \textbf{d)} shows the power spectral density $\psd_F(\omega, T) = \pi \hbar \spd(\omega) \coth{\big( \frac{\hbar \omega}{2 \kBT} \big )}$ of the stochastic noise. Choosing a Lorentzian form of the spectral density $\spd(\omega)$ gives rise to colored noise, leading to correlations, which are shown by the skewed distribution of the two-time correlation function in panel \textbf{c)}.}
    \label{fig:noise}
\end{figure}

To illustrate how this is done, let us begin by recalling the centerpiece of non-equilibrium statistical physics: classical Brownian motion, i.e. the seemingly random motion of a classical particle in a medium \cite{brown1828, einstein1905, vonsmoluchowski1906, langevin1908, perrin1909, kubo_book, zwanzig_book}. A Brownian particle trapped in a potential $V(x)$ can be effectively described as a stochastic process through a Langevin equation \cite{langevin1908, fokker1914, planck1917, vankampen_book}
\begin{equation} \label{eq:cl_langevin}
    m \, \ddot{x}(t) + \gamma \, \dot{x}(t) + V(x) = F(t),
\end{equation}
where $m$  and $x$ are the mass and position of the particle, respectively, and $\gamma$ is the damping parameter. Here, $F(t)$ describes the force from the environment at a given time $t$, which is taken to be stochastic, i.e. it is a random variable with an associated probability distribution \cite{sekimoto_book}. Fig.~\ref{fig:noise} illustrates this: panel a) shows a single realization of the stochastic force $F$ as a function of time $t$, b) shows the probability distribution of the stochastic force, which is taken to be Gaussian, c) its autocorrelation function, and d) its power spectrum in frequency space. Crucially, the stochastic force and the dissipation, arising from the damping term $\gamma \dot{x}$, are linked through the fluctuation dissipation theorem (FDR) \cite{callen1951, kubo1966}, $\expval{F(t) F(t')}_\sto = 2 \kBT \, \gamma \, \delta(t - t')/m$, where $k_{\mathrm{B}}$ is the Boltzmann constant, $T$ the temperature, and $\expval{}_\sto$ indicates the average over different realizations of the stochastic force. In standard stochastic methods, the FDR is used to completely capture the influence of the environment on the system dynamics \cite{vankampen_book, sekimoto_book}. Here, the force is uncorrelated in time, i.e. it is white noise \cite{sekimoto_book}. As we will see later, this noise can be colored to allow for memory effects in the system dynamics \cite{anders2022}.

Whilst this open system can be effectively solved classically with stochastic methods, solving the equivalent quantum equation - the `quantum Langevin equation' \cite{caldeira1983qbm} - is not so straightforward. Classically, knowledge about the environment is incorporated by considering the system of interest under the influence of a stochastic force \cite{seifert2012}. However, quantum mechanically, the classical variables are now operators that act on the full system + bath Hilbert space \cite{breuer_book, weiss_book}. This can lead to a build-up of entanglement between the system and its surroundings, which makes isolating the influence of the environment and finding the state of the system trickier.

The most commonly used approach to solve the open system dynamics is to assume the limit of weak coupling to the environment, where an effective description of the system alone can be obtained in the form of a master equation \cite{davies1974, gorini1976, lindblad1976}. This can then be treated stochastically as an average over quantum trajectories \cite{gisin1992, plenio1998, percival_book}. However, such derivations make several strong assumptions, such as the Born-Markov approximation, which imposes Markovian (memoryless) dynamics \cite{breuer_book}. To go beyond the weak-coupling regime and account for memory effects, one generally has to resort to computationally demanding numerical methods, such as TEMPO \cite{strathearn2018, gribben2022}, HEOM \cite{tanimura1989, jin2008}, TEDOPA \cite{prior2010, chin2010} and the RC mapping technique \cite{strasberg2016, nazir2018}. Therefore, effective approaches that are able to capture key quantum features without the need to solve the full quantum model, analogous to that used to study classical Brownian motion, are highly desirable.

One may ask whether we can model the quantum dynamics by coloring the noise in the classical Langevin equation \eqref{eq:cl_langevin} to have a `quantum' spectrum? In fact, this is what is known as the `quasiclassical' Langevin equation \cite{adelman1976, koch1980, schmid1982, caldeira1983qbm, grabert1984, eckern1990, kleinert1995, lu2012, stella2017}. For a class of models with a linear system and environmental coupling, e.g. the harmonic oscillator with bilinear environmental coupling, using stochastic noise that is Gaussian distributed (see Fig.~\ref{fig:noise}) reproduces exactly the moments in position and momentum of the fully-quantum harmonic oscillator \cite{schmid1982, grabert1984}. These moments can then be used to completely characterize the equilibrium state of the system \cite{schmid1982, grabert1984} and, for initial states that are also Gaussian, the dynamical state of the system \cite{ferraro2005}.

Whilst this case has been well explored, and the correspondence of the quasiclassical and open quantum moments established \cite{schmid1982, grabert1984}, to our knowledge, there is no single document that provides, in short, all the details of the open systems approach as well as those of the stochastic technique. This manuscript fills this gap: it brings together analytical expressions for covariances of the open quantum harmonic oscillator in time as well as in the steady state, and demonstrates why they are matched by a quasiclassical stochastic approach. This will provide a foundation to explore how other non-linear open quantum
systems, such as spins in magnetic materials \cite{barker2019, spidy, berritta2024}, may be modelled with similar classical stochastic methods.

The paper is structured as follows: in Section~\ref{sec:oqs_approach}, we solve for the dynamical state of a damped quantum harmonic oscillator using the standard
open quantum systems approach (OQu) and also the equivalent \emph{classical} open systems approach (OCl). In Section~\ref{sec:sto_approach}, we illustrate how
one can use a quasiclassical stochastic approach (StoQu), i.e. a classical equation of motion with colored `quantum' noise, to recover the OQu dynamical state for
arbitrary coupling to the environment \cite{grabert1984}. We also show the agreement of the stochastic classical (StoCl) dynamical state, i.e. classical equation of
motion with classical white noise, with the OCl dynamical state \cite{ford1965, zwanzig1973, seifert2012, sekimoto_book}. In Section~\ref{sec:mean_force}, we
discuss when one can and cannot model the steady state as the standard Gibbs state. Finally, in Section~\ref{sec:network}, we give an overview of how one can
employ stochastic techniques with quantum-like noise to model quantum heat transport in harmonic networks \cite{dhar2001, dhar2006, wang2007, wang2008, wang2009,
segal2016, sevincli2019, li2021}.

\section{Open Quantum Systems Approach} \label{sec:oqs_approach}

We will first recap the standard open quantum systems (OQu) treatment. Our aim is to solve for the state of a quantum harmonic oscillator in contact
with an environment consisting of a continuum of quantum harmonic oscillators: a paradigmatic model of an open quantum system \cite{caldeira1983qbm,
caldeira1983quantum, weiss_book, breuer_book, hanggi2005, lampo_book}. As we will see, for linear coupling to the environment, the total Hamiltonian is
quadratic and therefore generates \emph{Gaussianity-preserving} dynamics \cite{ferraro2005}. This means that the steady states and, for a broad class of
initial states that are Gaussian, the dynamical states can be completely characterized by the first and second moments in position and momentum, i.e. the mean
and covariance, see Eqs.~\eqref{eq:means}~-~\eqref{eq:I}.

The system Hamiltonian is that of the standard quantum harmonic oscillator
\begin{equation} \label{eq:sys_hamiltonian}
    \op{H}_\sys = \frac{1}{2} \left[ \frac{\op{p}^2}{m} + m \wHO^2 \, \op{x}^2 \right],
\end{equation}
where the position and momentum operators $\op{x}$ and $\op{p}$ satisfy the canonical commutation relations $\left[ \op{x}, \op{p}\right] = i \hbar$. Here, and in what follows, boldface symbols will denote operators. We will further model the environment as an infinite collection of harmonic oscillators at all frequencies $\omega$, so that the bare bath Hamiltonian is
\begin{equation} \label{eq:bath_hamiltonian}
    \op{H}_\bath = \frac{1}{2} \int_0^\infty \dd\omega \left[ \frac{\op{p}_\omega^2}{m_\omega} + m_\omega \omega^2 \op{x}_\omega^2 \right],
\end{equation}
with $\left[ \op{x}_\omega,\op{p}_{\omega'} \right] = i \hbar \delta(\omega - \omega')$. This approach is particularly well suited to describe the physics of systems where the main contribution to dissipation is due to bosonic modes of the environment, e.g. due to an electromagnetic field or vibrational modes of an underpinning lattice.

Finally, we consider that the system and bath interact via a bilinear interaction. Here, we choose to incorporate it so that the bath Hamiltonian plus interaction reads
\begin{equation} \label{eq:int_hamiltonian}
    \op{H}_{\bath + \inter} =  \int_0^\infty \dd\omega \, \frac{1}{2 m_\omega} \left[ \op{p}_\omega^2 + \left( m_\omega \omega \, \op{x}_\omega + \frac{c_\omega}{\omega} \op{x} \right)^2 \right],
\end{equation}
where $c_\omega$ is the coupling strength to the bath mode of frequency $\omega$. Expanding the square in $\op{H}_{\bath + \inter}$ gives the bare bath Hamiltonian \eqref{eq:bath_hamiltonian}, plus the desired linear interaction term proportional to $\op{x} \, \op{x}_\omega$, and finally a term dependent only on the system operators and proportional to $\op{x}^2$. This last term is known as the \emph{counter-term}, and much discussion exists in the literature about the need to account for it \cite{caldeira1983quantum, correa2023}. Here, we include it, which is physically well-motivated for cases where, for example, the environment represents electromagnetic fields \cite{caldeira1983quantum}. To quantify the effects of the bath on the system, the \emph{spectral density}, defined as
\begin{equation} \label{eq:spectral_density}
    \spd(\omega) = \frac{c_\omega^2}{2 m_\omega \omega},
\end{equation}
will be a useful quantity. With the total Hamiltonian $\op{H}_\tot = \op{H}_\sys + \op{H}_{\bath + \inter}$ defined, it is straightforward to write the Heisenberg equations of motion for the system and bath operators $\op{x}$, $\op{p}$, $\op{x}_\omega$, and $\op{p}_\omega$. From these, as we show in detail in Appendix~\ref{sec:quantum_dynamics_appendix}, one can find a compact integro-differential equation for the system operators, known as the `quantum Langevin equation' \cite{caldeira1983qbm}, given by
\begin{equation} \label{eq:qu_langevin}
    m \, \ddot{\op{x}}(t) + m \wHOS^2 \, \op{x}(t) - \int_0^t \dd t' \, \memker(t - t') \, \op{x}(t') = \op{F}(t),
\end{equation}
where the frequency is renormalized as $\wHOS^2 = \wHO^2 + \frac{2}{m} \int_0^\infty \dd\omega \, \frac{\spd(\omega)}{\omega}$ due to the counter-term. A dissipation kernel $\memker(\tau)$ appears in the above equation, and is given by \cite{breuer_book}
\begin{equation} \label{eq:memory_kernel}
    \memker(\tau) = 2 \, \hbar \, \Theta{(\tau)} \int_0^\infty \dd\omega \, \spd(\omega) \sin(\omega\tau),
\end{equation}
with $\Theta{(\tau)}$ the Heaviside step function. This kernel introduces memory (non-Markovian effects) into the system dynamics. This can be thought of as follows: the system interacts with the bath at $t'$, the bath then interacts with the system again at $t$, which means that the state of the system at $t$ depends on the system state at previous time $t'$. The quantum force exerted by the environment, $\op{F}(t)$, is given by
\begin{equation} \label{eq:qu_force}
    \op{F}(t) = - \int_0^\infty \dd\omega \, c_\omega \left[ \op{x}_\omega(0) \cos(\omega t) + \frac{\op{p}_\omega(0)}{m_\omega \omega} \sin(\omega t) \right],
\end{equation}
and accounts for the system dynamics' dependence on the state of the bath, where $\op{x}_\omega(0)$ and $\op{p}_\omega(0)$ are the initial operators for the bath position and momentum.

We can see that the quantum Langevin equation \eqref{eq:qu_langevin} takes the same form as the classical Langevin equation \eqref{eq:cl_langevin}, but where the classical variables and stochastic force term have been replaced with quantum operators. Here, we have chosen a specific form of a quadratic
potential $V(x)$ and have generalized it to include a dissipation kernel, but this can also be included in the classical equivalent, see
Eq.~\eqref{eq:sto_langevin}. For an Ohmic spectral density $\spd_\mathrm{Ohm}(\omega) = \gamma \omega/\pi$, the integration of the dissipation kernel
\eqref{eq:memory_kernel} gives rise to the damping term $- \gamma \, \dot{\op{x}}$, see Ref.~\onlinecite{hanggi2005}, which appears in \eqref{eq:cl_langevin}. Here, the system state has no
memory and the dynamics at time $t$ is `Markovian', i.e. the state of the system at time $t$ is independent of the state of the system at previous times $t'$.

For the following discussion, we will consider a general spectral density $\spd(\omega)$. It will be useful to introduce the noise autocorrelation function \cite{breuer_book}
\begin{equation} \label{eq:noise_autocorr}
    \noiseker(\tau) = \expval{\{ \op{F}(t), \, \op{F}(t - \tau) \}}
\end{equation}
where $\{\cdot, \cdot\}$ is the anti-commutator and $\expval{\cdot} = \tr[ \, \cdot \, \op{\rho}_0 ]$ is the quantum expectation value over the initial state $\op{\rho}_{\sysbath}(0)$ of the system and bath. Here, we take the initial state to be uncorrelated, i.e. $\op{\rho}_{\sysbath}(0) = \op{\rho}_S \otimes \op{\tau}_\bath$, where $\op{\tau}_\bath$ is the Gibbs state of the environment alone at temperature $T$
\begin{equation} \label{eq:gibbs_bath}
    \op{\tau}_\bath(T) = \frac{1}{Z_\bath} e^{-\op H_\bath/\kBT},
\end{equation}
where $Z_\bath = \tr \left[ e^{- \op{H}_\bath/\kBT} \right]$ is the partition function of the bath. Explicitly evaluating \eqref{eq:noise_autocorr} gives
\begin{equation} \label{eq:noise_autocorr_2}
    \noiseker(\tau) = 2 \int_0^\infty \dd\omega \frac{\spd(\omega)}{\omega} \noiseQu(\omega, T) \cos(\omega \tau),
\end{equation}
where the `quantum noise' spectrum $\noiseQu(\omega, T)$ is given by
\begin{equation} \label{eq:qu_noise}
    \noiseQu(\omega, T) = \hbar \omega \coth \left( \frac{\hbar\omega}{2\kBT} \right).
\end{equation}
Note that, here, the $\coth$ functional dependence of temperature arises from the thermal average of the bath's quantum harmonic oscillators.

The dynamical moments of the system oscillator position and momentum can be found by applying a Laplace transform to \eqref{eq:qu_langevin} - see Appendix.~\ref{sec:quantum_dynamics_appendix}. Doing so gives the mean position and momentum of the oscillator $\mu_{q}(t) = \expval{\op{q}(t)}, \, q \in \{x, p \}$ as \cite{correa2023}
\begin{subequations} \label{eq:means}
    \begin{align}
        \mu_x(t) &= g_1(t) \, \mu_x(0) + g_2(t)\, \mu_p(0), \label{eq:mean_x_t} \\
        \mu_p(t) &= g_3(t) \, \mu_x(0) + g_1(t) \, \mu_p(0). \label{eq:mean_p_t}
    \end{align}
\end{subequations}
with
\begin{subequations} \label{eq:g}
    \begin{align}
        g_1(t) &= \invlap_t \left[ \frac{s}{s^2 + \wHOS^2 - \lap{\memker}(s)/m} \right], \label{eq:g1} \\
        g_2(t) &= \invlap_t \left[ \frac{1}{m(s^2 + \wHOS^2) - \lap{\memker}(s)} \right], \label{eq:g2} \\
        g_3(t) &= \invlap_t \left[ \frac{\lap{\memker}(s) - m \, \wHOS^2}{s^2 + \wHOS^2 - \lap{\memker}(s)/m} \right]. \label{eq:g3}
    \end{align}
\end{subequations}
Here, $\lap{f}(s) = \int_0^\infty \dd t \, e^{-s \, t} f(t)$ stands for Laplace transform and $\invlap_t \left[ \, g \, \right]$ the inverse transform. 

One can also explicitly write the covariances $\sigma_{qq'}(t) = \frac{1}{2} \expval{ \{\op{q}(t), \, \op{q}'(t) \} } - \expval{\op{q}(t)} \expval{\op{q}'(t)}$ as follows \cite{lampo_book}
\begin{widetext}
    \begin{subequations} \label{eq:covs}
        \begin{align}
            \sigma_{xx}(t) &= \int_0^\infty \dd\omega \, \frac{\spd(\omega)}{\omega} \, \noise(\omega, T) \, \Re{\left( \int_0^t \dd t_1 \, g_2(t_1) e^{-i \omega t_1} \int_0^t \dd t_2 \, g_2(t_2) e^{i \omega t_2} \right)} + I_{xx}(t), \label{eq:cov_xx} \\
            \sigma_{xp}(t) &= m \int_0^\infty \dd\omega \, \frac{\spd(\omega)}{\omega} \, \noise(\omega, T) \, \Re{\left( \int_0^t \dd t_1 \, g_2(t_1) e^{-i \omega t_1} \int_0^t \dd t_2 \, g_1(t_2) e^{i \omega t_2} \right)} + I_{xp}(t), \label{eq:cov_xp} \\
            \sigma_{pp}(t) &= m^2 \int_0^\infty \dd\omega \, \frac{\spd(\omega)}{\omega} \, \noise(\omega, T) \, \Re{\left( \int_0^t \dd t_1 \, g_1(t_1) e^{-i \omega t_1} \int_0^t \dd t_2 \, g_1(t_2) e^{i \omega t_2} \right)} + I_{pp}(t), \label{eq:cov_pp}
        \end{align}
    \end{subequations}
\end{widetext}
with the noise term $N(\omega, T) = \noiseQu(\omega, T)$. Here, the $g_j$ are given in \eqref{eq:g} and the $I_{qq'}$ are given by
\begin{align} \label{eq:I}
    \begin{pmatrix}
        I_{xx}(t) \\
        I_{xp}(t) \\
        I_{pp}(t)
    \end{pmatrix} =
    \begin{pmatrix}
        g_1^2 & 2 g_1 g_2 & g_2^2 \\
        g_1 g_3 & g_1^2 + g_2 g_3 & g_1 g_2 \\
        g_3^2 & 2 g_1 g_3 & g_1^2
    \end{pmatrix} 
    \begin{pmatrix}
        \sigma_{xx}(0) \\
        \sigma_{xp}(0) \\
        \sigma_{pp}(0)
    \end{pmatrix},
\end{align}
where the notation for the time dependence of the $g_j$ has been suppressed.

It is important to highlight that, for initial states that are Gaussian, the expressions \eqref{eq:g}~and~\eqref{eq:covs} fully solve the system state at any time $t$. This is true for any spectral density $\spd(\omega)$, and therefore for arbitrarily complex non-Markovian dynamics \cite{grabert1984}. That said, explicitly computing the inverse Laplace transforms in \eqref{eq:g} cannot be done analytically in general. However, a Lorentzian spectral density \cite{anders2022}
\begin{equation} \label{eq:spectral_density_lor}
    \spd_\mathrm{Lor}(\omega) = \frac{1}{\pi} \frac{\lambda^2 \Gamma \omega}{(\omega_0^2 - \omega^2)^2 + \Gamma^2 \omega^2},
\end{equation}
is one case with a known analytical solution. Here, $\lambda$ is the system-bath coupling strength, $\Gamma$ the peak width, and $\omega_0$ the resonant frequency. Furthermore, although being simple to analyze, it is worth noting that the Ohmic spectral density is not strictly physical, as it causes the spectral integrals to diverge \cite{spreng2015}. This includes quantities such as the renormalized frequency $\wHOS$. Therefore, it is necessary to introduce
some cutoff to the coupling to high frequencies or to use a more physically motivated functional form for $\spd(\omega)$. Moreover, the Ohmic spectral density
is not capable of capturing non-Markovian effects \cite{ceriotti2010, ceriotti2011, lu2019}, which can become relevant for experiments at the nanoscale and ultra-short timescales. As such, we will use a Lorentzian form $\spd_\mathrm{Lor}(\omega)$ for the numerical simulations.

\subsection{Classical open system} \label{sec:ocs_approach}

One can also apply the above analytical open systems approach fully classically (OCl), i.e. a classical oscillator coupled to a bath composed of classical oscillators. This yields the same dynamical moments as the quantum equivalent (see Appendix~\ref{sec:classical_dynamics_appendix} for details) but with the noise term $\noise(\omega, T)$ given by the classical noise
\begin{equation} \label{eq:cl_noise}
    \noiseCl(\omega, T) = 2 \kBT,
\end{equation}
in Eqs.~\eqref{eq:covs}. In particular, in the long-time limit, averaging over the bath degrees of freedom, one finds that the system settles into the canonical equilibrium state, $\op{\rho}_{\mathrm{eq}} \propto e^{- \beta H_\sys}$, regardless of the strength of the coupling, which is explored in Appendix.~\ref{sec:exact_appendix}.

\medskip

Here, we presented in brief the steps of the standard open systems approach for finding the moments \eqref{eq:means}~and~\eqref{eq:covs} of a dissipative quantum harmonic oscillator coupled linearly to its environment \cite{breuer_book, correa2023}. For Gaussian initial states, one can use these moments to fully characterize the dynamical state of the system \cite{ferraro2005}. We also considered the equivalent \emph{classical} system, noting that the only change to the moments is the form of the noise term $\noise(\omega, T)$. We will now discuss the link between the environmental force $\op{F}(t)$ and the dissipation kernel $\memker(t)$.

\subsection{Fluctuation-dissipation relation}

The environmental force $\op{F}(t)$ and the dissipative kernel $\memker(t)$ can be shown to satisfy the quantum fluctuation-dissipation relation (FDR) \cite{kubo1966}. More concretely, let $\psd_F(\omega)$ be the power spectral density of the environment force $\op{F}(t)$, that is \cite{anders2022}
\begin{equation} \label{eq:psd}
    \psd^\qu_F(\omega) = \frac{1}{2} \int_{-\infty}^{\infty} \dd\tau \expval{\{\op{F}(t), \, \op{F}(t-\tau)\}} \, e^{i \omega \tau}.
\end{equation}
Substituting in the noise autocorrelation function \eqref{eq:noise_autocorr_2} yields
\begin{equation} \label{eq:qu_FDR_2}
    \psd^\qu_F(\omega, T) = \pi \frac{\spd(\omega)}{\omega} \noiseQu(\omega).
\end{equation}
Furthermore, since $\Im{ \left[ \tilde{\memker}(\omega) \right]} = \pi \, \spd(\omega)$ (see Ref.~\onlinecite{anders2022}), using the definition of the quantum noise \eqref{eq:qu_noise} allows us to express Eq.~\eqref{eq:qu_FDR_2} as
\begin{equation} \label{eq:qu_FDR}
    \psd^\qu_F(\omega, T) = \hbar\,\Im{\left[ \tilde{\memker}(\omega) \right] \coth \left( \frac{\hbar\omega}{2\kBT} \right)},
\end{equation}
with $\tilde{\memker}(\omega)$ the Fourier transform of the memory kernel \eqref{eq:memory_kernel}. In analogy with the classical FDR (see introduction), the \emph{quantum} FDR \eqref{eq:qu_FDR} relates the power spectral density of the environment force $\op{F}(t)$ with the dissipation kernel $\memker(t)$.

From \eqref{eq:qu_FDR_2}, we can clearly see that there are two contributions to the force power spectral density, which have very different physical origins. The first term, $\spd(\omega)$, comes from the specific form of the coupling of the system to the environment. On the other hand, the second term, $\noiseQu(\omega, T)$, does not depend on the system but exclusively on the bath. It fully encodes the temperature dependence of the environment force $\op{F}(t)$ and is the only place where the quantum signature in the power spectral density manifests itself. Indeed, at high temperatures, i.e. $\kBT \gg \hbar\omega$ for all relevant frequencies, it becomes $\noiseQu(\omega, T) \approx 2\kBT = \noiseCl(\omega, T)$, which is nothing more than the classical temperature dependence of the noise on a Brownian particle \cite{blundell_book}.

Despite involving operators, Eq.~\eqref{eq:psd} looks like a power spectrum of a fluctuating force. In the following section, we will see that the effect of the environment on the dynamical state of the system can indeed be captured by a stochastic process \cite{grabert1984}.

\section{Stochastic approach} \label{sec:sto_approach}

It is well known that the stochastic approach with classical noise (StoCl) will reproduce the dynamical state of the open classical (OCl) \cite{ford1965, zwanzig1973, seifert2012, sekimoto_book}. This agreement is guaranteed by imposing that the fluctuations of the stochastic force obey the classical FDR. A question of great practical interest is if one can modify the stochastic force statistics such that the dynamics reproduces the behaviour of the full open quantum system? This question gave rise to the development of quasiclassical Langevin equations \cite{adelman1976, koch1980, schmid1982, caldeira1983qbm, grabert1984, eckern1990, kleinert1995, lu2012, stella2017}, where the classical stochastic noise is colored to match the quantum spectrum \eqref{eq:qu_noise}. For the model considered here, with stochastic noise that is Gaussian-distributed, the moments in position and momentum from the quasiclassical Langevin equation match exactly those from the OQu approach \cite{schmid1982, grabert1984}. This then allows one to solve for the equilibrium state of the system for arbitrary forms and strengths of environmental coupling \cite{grabert1984}. Furthermore, for Gaussian initial states, the moments can also be used to characterize the dynamical system state \cite{ferraro2005}. Even in some anharmonic setups, such as those encountered in molecular dynamics (MD) simulations, this quasiclassical approach can yield highly accurate results \cite{ceriotti2009, dammak2009, ceriotti2010, ceriotti2011, ceriotti2016, lu2019}. Another application of this approach has been in the context of modelling quantum heat transport \cite{dhar2001, dhar2006, wang2007, wang2008, wang2009, segal2016, sevincli2019, li2021}. Similar methods involving coloring the classical noise to have a `quantum' spectrum have also been employed in quantum electrodynamics \cite{rytov_book, boyer1975, henkel2016}, in the modelling of magnetic materials \cite{barker2019, anders2022, berritta2024, spidy}, and further afield in the study of many-body decoherence \cite{chenu2017}. Despite advances allowing for the inclusion of non-linearities in the quasiclassical Langevin equation, we focus here on a linear system (harmonic oscillator) coupled linearly to its environment.

We will now explicitly calculate $\mu_q$ and $\sigma_{q q'}$ using the quasiclassical stochastic treatment, i.e. a classical particle subjected to stochastic quantum noise (StoQu). To begin, we replace the operators $\op{x}(t)$ and $\op{F}(t)$ in the quantum Langevin equation \eqref{eq:qu_langevin} with the real classical variables $x(t)$ and $F(t)$ to obtain
\begin{equation} \label{eq:sto_langevin}
    m \, \ddot{x}(t) + m \, \wHOS^2 \, x(t) - u(t) = F(t),
\end{equation}
where $u(t) = \int_0^t \dd t' \, \memker(t - t') \, x(t')$ encodes the dissipation \cite{anders2022}. Since $F(t)$ is now a classical variable, it commutes with itself at all other times. As such, the definition of the power spectral density of the stochastic field \eqref{eq:psd} becomes
\begin{equation} \label{eq:sto_psd}
    \psd_F^\sto(\omega) = \int_{-\infty}^{\infty} \dd\tau \expval{F(t) F(t - \tau)}_\sto e^{i \omega \tau},
\end{equation}
where $\expval{}_\sto$ indicates an average over stochastic trajectories. For the fully classical model, i.e. a classical particle subjected to classical noise (StoCl), the FDR takes the form
\begin{equation} \label{eq:sto_cl_FDR}
    \psd_F^{\sto, \cl}(\omega, T) = \pi \frac{\spd(\omega)}{\omega} \noise^\cl(\omega, T).
\end{equation}
However, in the StoQu approach, we instead impose that $F(t)$ obeys the FDR with quantum noise \cite{schmid1982, grabert1984}
\begin{equation} \label{eq:sto_qu_FDR}
    \psd_F^{\sto, \qu}(\omega, T) = \pi \frac{\spd(\omega)}{\omega} \noiseQu(\omega, T).
\end{equation}
Eq.~\eqref{eq:sto_langevin} with power spectral density \eqref{eq:sto_qu_FDR} yields a quasiclassical Langevin equation, i.e. we are solving the classical Langevin equation but with a `quantum' form of the environmental noise.

Here, we also only need to consider the mean and covariance to completely characterize the state of the system if we assume the noise has a Gaussian distribution, as illustrated in Fig.~\ref{fig:noise}. This is because the solution of a linear stochastic differential equation will also always be Gaussian
and hence can be described completely with the first and second moments \cite{sobczyk_book}. It is worth noting that the requirement for the noise to be
Gaussian distributed does not mean that the noise cannot be colored, which allows for the inclusion of memory effects in the system - this is illustrated in
Fig.~\ref{fig:noise}.

We can solve \eqref{eq:sto_langevin} by applying a Laplace transform to 
\begin{subequations} \label{eq:sto_eqm}
    \begin{align}
        \dot{x}(t) &= \frac{p(t)}{m}, \label{eq:sto_eqm_1} \\
        \dot{p}(t) &= - m \, \wHOS^2 x(t) + F(t) + \int_0^t \dd t' \, \memker(t - t') \, x(t'), \label{eq:sto_eqm_2}
    \end{align}
\end{subequations}
and rearranging to give expressions for $x(t)$ and $p(t)$. The first moments $\mu^\sto_q(t) = \llangle q(t) \rrangle_0, \, q \in \{ x, p\}$ can then be obtained by taking both a stochastic average and an average over initial conditions $\expval{}_0$. For simplicity, when we take both averages $\langle \langle \rangle_\sto \rangle_0$, we will instead just write $\llangle \rrangle_0$. Note that the averaging over initial conditions is required to compare to the fully quantum model since, quantum mechanically, the initial state necessarily has uncertainty. Evaluating the first moments in the stochastic framework gives exactly Eqs.~\eqref{eq:means} from the OQu treatment. With the expressions for the first moments \eqref{eq:means}, one can find the second moments in the StoQu approach as $\sigma_{qq'}^\sto(t) = \llangle q(t) \, q'(t) \rrangle_0 - \llangle q(t) \rrangle_0 \llangle q'(t) \rrangle_0$. In fact, these moments agree exactly with those from the OQu approach Eqs.~\eqref{eq:covs} with quantum noise $\noise(\omega, T) = \noiseQu(\omega, T)$, with further details presented in Appendix~\ref{sec:stochastic_dynamics_appendix}.

It is worth noting that, in the literature, Ehrenfest's theorem has also been used directly to show this agreement: the linearity of the problem means the quantum expectation values will follow the classical trajectories \cite{rae_book, grabert1984}. Furthermore, the linearity means that both the classical and quantum response functions coincide \cite{caldeira1983quantum, grabert1984, grabert1988, hanggi2005}. Therefore, since the response function is related to the position autocorrelation function, it can also be used to show the agreement of the second moments \cite{grabert1984}, as we have done here by explicitly calculating the moments using both the open systems and quasiclassical approaches.

\begin{figure}
    \centering
    \includegraphics{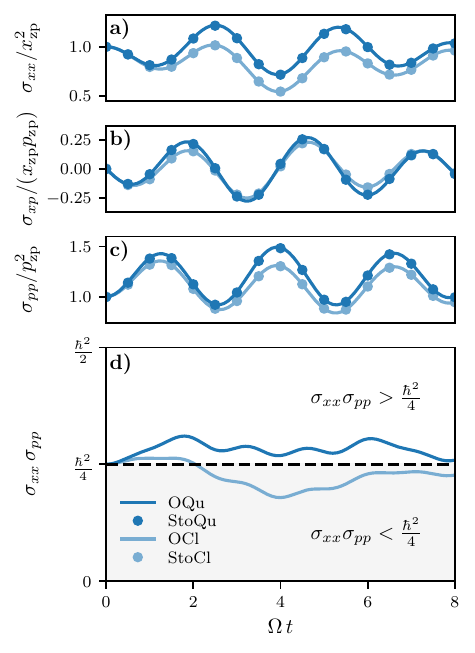}
    \caption{Panels \textbf{a)} - \textbf{c)}: agreement of the dynamical covariances $\sigma_{xx}$(t), $\sigma_{xp}(t)$ and $\sigma_{pp}(t)$, see Eqs.~\eqref{eq:covs}~and~\eqref{eq:I}, from the OQu (dark blue lines) and StoQu (dark blue dots) approaches. Also plotted are the dynamical covariances from the OCl approach (light blue lines) and the StoCl approach (light blue dots). Panel \textbf{d)} shows the violation of the uncertainty relation by the classical dynamics, indicated by the drop of the light-blue line into the grey region. In panels $\mathbf{a)}$ - $\mathbf{c)}$, the variances have been normalised by the zero-point magnitudes $x_{\rm{zpm}} = \sqrt{\hbar / (2 m \wHO)}$ and $p_{\rm{zpm}} = \sqrt{(\hbar m \wHO)/2}$, and the Lorentzian parameters are $\lambda = 0.3 \, \lambdaUnits$, $\wLor = 0.5 \, \wHO$ and $\GammaLor = 0.1 \, \wHO$, and $T = 0.1 \hbar \wHO/\kB$. The initial variances are $\sigma_{xx}(0) = \sigma_{pp}(0) = 0.5$ and $\sigma_{xp}(0) = 0$.}
    \label{fig:dynamics} 
\end{figure}

As we have seen, the linearity of the problem allows one to show the agreement of the two methods analytically \cite{grabert1984}. Additionally, we can also evaluate the stochastic covariances numerically with \texttt{SpiDy.jl}: a dedicated Julia module for the stochastic simulation of open systems of spins and harmonic oscillators \cite{spidy}. Whilst not strictly necessary here, such tools will be required for non-linear setups, such as spin systems, where analytical solutions are currently unknown \cite{barker2019, spidy, berritta2024}. We employ the numerical stochastic simulation here for the exactly solvable case of the harmonic oscillator to highlight its capability. We illustrate the dynamical agreement of the analytical OQu (dark blue lines) and the numerically solved StoQu (dark blue dots) second moments (Eqs.~\eqref{eq:covs} with quantum noise $\noise(\omega, T) = \noiseQu(\omega, T)$) as a function of time $t$ in Fig.~\ref{fig:dynamics}. Additionally, we also highlight the classical agreement by plotting the dynamical second moments from OCl (light blue lines) and StoCl (light blue dots) treatments, given by \eqref{eq:covs} but with a classical noise term $\noiseCl(\omega, T)$. Of particular interest is that the stochastic approach with quantum noise obeys the uncertainty relation, as shown in panel d), whilst the fully classical dynamics does not.

It is worth noting that the covariances \eqref{eq:covs}~-~\eqref{eq:I} are obtained when using the \emph{symmetrized} quantum correlation function
\eqref{eq:psd}. Not performing the symmetrization in \eqref{eq:psd} gives an imaginary component that cannot be simulated classically \cite{wang2007}. Whilst the single-time covariances are fully captured in the quasiclassical approach, the quasiclassical treatment is not the same as the full open systems description \cite{hanggi2005}, i.e. there is a difference in the \emph{unsymmetrized} two-time correlation functions. Nevertheless, the quasiclassical approach can still be used to capture the open quantum dynamical state of the quantum harmonic
oscillator coupled linearly to its environment \cite{schmid1982, grabert1984}.

In this section, we illustrated the agreement of the quasiclassical stochastic dynamical moments \eqref{eq:means} and \eqref{eq:covs} for a dissipative harmonic oscillator with those from the full open quantum systems approach \cite{schmid1982, grabert1984}. Importantly, this is true for arbitrary spectral densities and environmental coupling strengths \cite{grabert1984}. For Gaussian initial states, these moments can be used to completely characterize the dynamical state of the system \cite{ferraro2005}. Despite having an analytical form of these moments, here we solve the stochastic case numerically with the Julia module \texttt{SpiDy.jl}\cite{spidy}. The exact agreement provides a good testbed for such tools, which have already been employed to study non-linear setups where there exists no such analytical form, such as spin systems \cite{barker2019, anders2022}.

\section{Steady state and mean force state} \label{sec:mean_force}

In the previous section, we illustrated how one can use quasiclassical stochastic methods to evaluate the quantum dynamical state of a dissipative harmonic oscillator \cite{schmid1982, grabert1984}. We will now discuss the agreement of the two approaches in the steady state, specifically illustrating how the quasiclassical method recovers the quantum mean force Gibbs state.

\begin{figure}[t]
    \centering
    \includegraphics{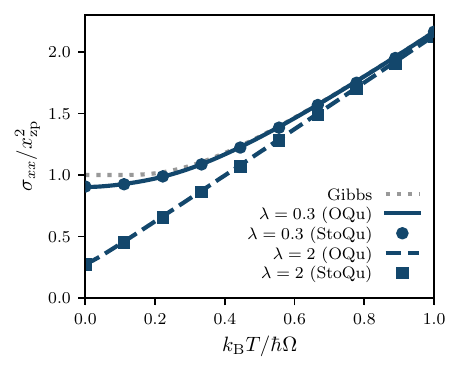}
    \caption{Agreement of the steady-state covariance $\sigma_{xx}(\infty)$, see Eq.~\eqref{eq:cov_xx_ss}, from the OQu (lines) and StoQu (markers) approaches for environmental coupling strengths of $\lambda = 0.3 \, \lambdaUnits$ (solid line, circle markers) and $\lambda = 2 \, \lambdaUnits$ (dashed line, square markers). Also plotted is $\sigma_{xx}$ calculated with the Gibbs state \eqref{eq:gibbs} (grey dotted line). One can see that increasing $\lambda$ consistently leads to a deviation of the steady state from the Gibbs state. Here, the variances have been normalised by the zero-point magnitude $x_{\rm{zpm}} = \sqrt{\hbar / (2 m \wHO)}$, and the other Lorentzian parameters are $\wLor = 0.5 \, \wHO$ and $\GammaLor = 0.1 \, \wHO$, and $T = 0.1 \hbar \wHO/\kB$. The initial variances are $\sigma_{xx}(0) = \sigma_{pp}(0) = 0.5$ and $\sigma_{xp}(0) = 0$.}
    \label{fig:steady} 
\end{figure}

In the long-time limit, the means \eqref{eq:means} vanish, and the covariances \eqref{eq:covs} become \cite{correa2023}
\begin{subequations} \label{eq:covs_ss}
    \begin{align}
        \sigma_{xx}(\infty) &= \int_0^\infty \dd\omega \, \frac{\spd(\omega)}{\omega} \, \noise(\omega, T) \, \abs{\lap{g}_2(i \omega )}^2, \label{eq:cov_xx_ss} \\
        \sigma_{xp}(\infty) &= 0, \label{eq:cov_xp_ss} \\
        \sigma_{pp}(\infty) &= m^2 \int_0^\infty \dd\omega \, \omega^2 \, \frac{\spd(\omega)}{\omega} \, \noise(\omega, T) \, \abs{\lap{g}_2(i \omega )}^2, \label{eq:cov_pp_ss}
    \end{align}
\end{subequations}
with $\noise(\omega, T) = \noiseQu(\omega, T)$ for the OQu and StoQu approaches, and $\noise(\omega, T) = \noiseCl(\omega, T)$ for the OCl and StoCl approaches (details in Appendix~\ref{sec:steady_state_appendix}). As an example, we show the agreement of the $\sigma_{xx}$ covariance in the steady state in Fig.~\ref{fig:steady} as a function of temperature, $T$. The moments are evaluated for two coupling strengths: $\lambda = 0.3 \, \lambdaUnits$ shown by the solid line (OQu) and dots (StoQu), and a stronger coupling of $\lambda = 2 \, \lambdaUnits$ shown by the dashed line (OQu) and squares (StoQu).

One might expect the steady state to match the Gibbs state of the harmonic oscillator, given by
\begin{equation} \label{eq:gibbs}
    \Gibbs = \frac{1}{Z_0} e^{-\beta \HHO} = \frac{1}{Z_0} \sum_{n=0}^\infty e^{-(n+1/2) \beta \hbar \wHO} \dyad{n},
\end{equation}
where $Z_0 = \tr[e^{-\beta \HHO}] = e^{-\beta \hbar \wHO / 2}/(1 - e^{-\beta \hbar \wHO})$ is the partition function. In Fig.~\ref{fig:steady}, we also plot $\sigma_{xx}$ calculated using the Gibbs state as a function of temperature (grey dotted line). We can see that the steady-state moments are well-approximated by the moments of the Gibbs state for very low coupling $\lambda = 0.3 \, \lambdaUnits$. However, at larger coupling $\lambda = 2 \, \lambdaUnits$, significant deviations appear. Indeed, whilst it is commonly assumed that the steady state of a system in contact with a thermal environment is the Gibbs state, this is only true in the limit of vanishingly small coupling \cite{trushechkin2022}. As soon as the coupling to the environment becomes non-negligible, one must account for the environment-induced corrections.

A natural question that then arises is if there is a generalization of the Gibbs state that might be valid at all coupling strengths. This is of
not only theoretical interest but also practical, particularly for problems beyond the harmonic oscillator that cannot be solved analytically where one must
resort to numerical methods. Indeed, in such a case, having a method to directly compute the steady state without needing to solve the long-time dynamics
becomes highly relevant. Within the community of strong coupling thermodynamics, it is commonly postulated that the correct steady state is given by the so-called \emph{mean force} Gibbs state \cite{miller2018, thingna2012, subasi2012, cresser2021, merkli2022_1, merkli2022_2, trushechkin2022, cerisola2024}, given by
\begin{equation} \label{eq:MF}
    \MF = \tr_B \left[ \frac{1}{Z} e^{-\beta \op{H}} \right].
\end{equation}
For the case of the dissipative harmonic oscillator, the mean force state has been calculated explicitly \cite{trushechkin2022, grabert1984, grabert1988, hanggi2005}. In this mean force framework, the steady-state covariances are \cite{trushechkin2022, philbin2012}
\begin{subequations} \label{eq:covs_ss_mf}
    \begin{align}
        \sigma_{xx}(\infty) &= \frac{1}{\pi} \int_0^\infty \dd\omega \, \frac{1}{\omega} \noiseQu(\omega, T) \Im{\left[ G(\omega) \right]}, \label{eq:cov_xx_mf} \\
        \sigma_{xp}(\infty) &= 0, \label{eq:cov_xp_mf} \\
        \sigma_{pp}(\infty) &= m^2 \frac{1}{\pi} \int_0^\infty \dd\omega \, \omega \, \noiseQu(\omega, T) \Im{\left[ G(\omega) \right]}, \label{eq:cov_pp_mf}
    \end{align}
\end{subequations}
where $G(\omega) = -(m\omega^2 - \wHOS^2[m - \chi(\omega)])^{-1}$ and
\begin{equation}
    \wHOS^2 \chi(\omega) = \mathcal{P} \int_0^\infty \dd\xi \, \frac{2 \xi J(\xi)}{\xi^2 - \omega^2} + i \pi \spd(\omega).
\end{equation}
Here, $\mathcal{P}$ denotes the principal value integral. Since, $\pi \spd(\omega) = \Im{\left[ K(\omega) \right]}$ (see Ref. \onlinecite{anders2022}), we can use the Kramers-Kronig relations to identify $\wHOS^2 \chi(\omega)$ as simply $K(\omega)$. With this, we obtain exactly the open systems expressions, i.e. Eqs.~\eqref{eq:covs_ss} with quantum noise $\noise(\omega, T) = \noiseQu(\omega, T)$.

In this section, we discussed the deviation of the steady state of the dissipative harmonic oscillator with increasingly strong environmental coupling from the canonical Gibbs state. Whilst the Gibbs state is the thermal state of the system Hamiltonian alone, the mean force Gibbs state is that of the global system + environment Hamiltonian. This mean force state has been shown to be the true steady state of the dissipative harmonic oscillator with linear environmental coupling for a broad class of initial states \cite{trushechkin2022, grabert1988, fleming2011}. In this paper, we considered an initially uncorrelated state to find the open quantum dynamics, i.e. $\op{\rho}_{\sysbath}(0) = \op{\rho}_S \otimes \op{\tau}_\bath$. For this case, we have illustrated the agreement of approaches explicitly by a match of the open systems steady-state covariances \eqref{eq:covs_ss} with the mean force predictions \eqref{eq:covs_ss_mf}.

\section{Non-equilibrium networks} \label{sec:network}

So far, we have seen how the open quantum dynamical state of a single oscillator coupled to a single bath can be reproduced using a quasiclassical stochastic treatment \cite{schmid1982, grabert1984}. In this section, we will illustrate how quasiclassical stochastic approaches can also be used to effectively model networks of harmonic oscillators and heat transport in the quantum regime \cite{dhar2001, dhar2006, wang2007, wang2008, wang2009, segal2016, sevincli2019, li2021}. To do so, we will first set up the open system Hamiltonian for a network of harmonic oscillators, each linearly coupled to an independent bath of quantum harmonic oscillators and show that the resulting dynamics can still be reproduced in the quasiclassical stochastic treatment. We then see how baths at different temperatures induce heat transport and how this can be captured stochastically.

\subsection{Open systems approach}

For a system which is now a network of oscillators (see Fig.~\ref{fig:network}), the system Hamiltonian can be written as \cite{martinez2013, freitas2014, freitas2017}
\begin{equation} \label{eq:sys_hamiltonian_network}
    \op{H}_{\sys} = \frac{1}{2} \left( \vect{\op{P}}^\mathsf{T} \mat{M}^{-1} \vect{\op{P}} + \vect{\op{X}}^\mathsf{T} \mat{V} \, \vect{\op{X}} \right),
\end{equation}
where $\vect{\op{X}}$, $\vect{\op{P}}$ are vectors containing the position and momentum operators of the oscillators, $\mat{M}$, $\mat{V}$ are positive and symmetric matrices describing the masses and potential, and $\mathsf{T}$ denotes the transpose. The oscillator network linearly couples to an environment consisting of an independent bath for each system oscillator, each initially in a thermal state at temperature $T_\alpha$, and uncorrelated with each other. The bath + interaction Hamiltonian is given by
\begin{align} \label{eq:int_hamiltonian_network}
    \op{H}_{\bath + \inter} &= \sum_{\alpha, \gamma} \, \int_0^\infty \dd\omega \, \frac{1}{2 m_{\omega, \alpha}} \bigg[ \op{p}_{\omega, \alpha}^2 \nonumber \\
    & \qquad \qquad + \left( m_{\omega, \alpha} \omega \, \op{x}_{\omega, \alpha} + \frac{C_{\omega, \alpha \gamma}}{\omega}\op{X}_\gamma \right)^2 \bigg],
\end{align}
where $\op{x}_{\omega, \alpha}, \op{p}_{\omega, \alpha}$ are the position and momentum operators of the bath oscillators with mass $m_{\omega, \alpha}$ and frequency $\omega$ in the bath $\alpha$, coupling to the system oscillator $\gamma$ with strength $C_{\omega, \alpha \gamma}$. In Appendix~\ref{sec:network_appendix}, we give details on the resulting dynamics and show how it is reproduced stochastically.

Baths at different temperatures induce a heat flow through the network, where a positive value indicates heat flow into the network. We can evaluate the steady-state heat currents as (see Appendix~\ref{sec:heat_current_appendix} for details)
\begin{equation} \label{eq:qu_currents}
    \dot Q_\alpha = \tr \left[ \mat{\Pi}_\alpha \, \mat{\bar{V}} \, \mat{C}_{xp} \, \mat{M}^{-1}\right],
\end{equation}
where $\mat{\Pi}_\alpha$ is the projector to the oscillators which the bath $\alpha$ couples to, $\mat{\bar{V}}$ the potential matrix including counter-term and $\mat{C}_{xp} = \lim_{t \to \infty} \frac{1}{2}\langle \{\vect{\op X}(t),\vect{\op P}(t)^\mathsf{T}\} \rangle $. The trace, $\tr()$, is taken over matrices containing the components of the different oscillators, and $\langle \rangle$ is the quantum mechanical average. Equivalently, the heat currents can be calculated using the Landauer-Caroli formula \cite{wang2007, freitas2017}. Eq.~\eqref{eq:qu_currents} shows that the heat currents can be expressed via the symmetrized correlation functions. As the symmetrized correlation functions are correctly reproduced in the stochastic approach, they can also be used to calculate the heat currents. In the following, we give a more convenient way to express the currents in the stochastic approach.

\subsection{Stochastic approach}

We now turn to the stochastic expression for the heat currents. Instead of calculating the heat currents using  Eq.~\eqref{eq:qu_currents} by evaluating $\mat{C}_{xp}$ from the stochastic oscillator position and momentum, we can express the heat currents more naturally using the output of the stochastic simulation. Straightforward manipulation of \eqref{eq:qu_currents}, and using the fact that the quantum operators are now classical variables, gives (see Appendix~\ref{sec:heat_current_appendix} for details)
\begin{equation} \label{eq:sto_currents}
    \dot Q_\alpha(t) = \left\langle \left(\vect{F}_\alpha(t) + \vect{u}_\alpha(t) \right) \cdot \vect{\dot{X}}(t) \right\rangle,
\end{equation}
where $\vect{F}_\alpha(t)$ and $\vect{u}_\alpha(t)$ are the contributions to the force and memory kernel from the $\alpha^\mathrm{th}$ bath. Those variables are naturally calculated to simulate the dynamics of the system, for example, in the aforementioned \texttt{SpiDy.jl} module.

\begin{figure}[t]
    \centering
    \includegraphics{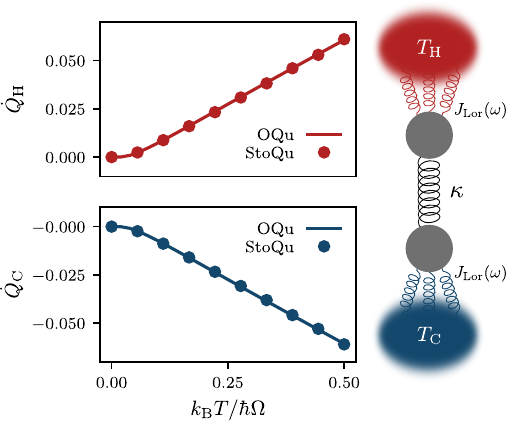}
    \caption{Agreement of the steady-state heat currents across a two oscillator network calculated using the open quantum systems approach (OQu, lines, \eqref{eq:qu_currents}) and the quasiclassical stochastic approach (StoQu, dots, \eqref{eq:sto_currents}) techniques. The oscillators are coupled together with strength $\kappa = 0.1 \hbar \wHO/m^2$ and the temperature gradient is $T_{\mathrm{H}} = 10 T_{\mathrm{C}}, T_{\mathrm{C}} = T$. As the system is in its steady state, the two heat currents have the same magnitude but opposite sign. The parameters that characterize the Lorentzian spectral density $\spd_\mathrm{Lor}(\omega)$ for each oscillator are $\lambda = 0.3 \, \lambdaUnits$, $\wLor = 0.5 \, \wHO$ and $\GammaLor = 0.8 \, \wHO$. For each oscillator, the initial variances are $\sigma_{xx}(0) = \sigma_{pp}(0) = 0.5$ and $\sigma_{xp}(0) = 0$.}
    \label{fig:network} 
\end{figure}

We illustrate the agreement of the heat currents in the steady state by considering a simple network of two oscillators, each in contact with their own bath at different temperatures $T_{\mathrm{H}} = 10\, T$ and $T_\mathrm{C} = T$. In Fig.~\ref{fig:network}, we plot both the open systems \eqref{eq:qu_currents} (OQu, lines) and stochastic \eqref{eq:sto_currents} (StoQu, dots) heat currents in the non-equilibrium steady state as a function of $T$. This shows that the quasiclassical stochastic approach can not just capture the open quantum system dynamics of a harmonic oscillator - or a network of harmonic oscillators - but also model heat transport in the quantum regime. Indeed, this has been readily exploited to calculate heat transport, e.g. in nanostructures \cite{wang2008} or molecular junctions \cite{segal2016}.

\section{Conclusion}

In this paper, we have given an overview of the use of quasiclassical stochastic methods for the modelling of open quantum systems, bringing all relevant results together in one place. In particular, we have seen how, for the case of a dissipative harmonic oscillator, a quasiclassical stochastic approach can recover the first and second moments of the open quantum system for arbitrary coupling to the environment \cite{grabert1984}. This allows one to completely characterize the steady state and, for Gaussian initial states - as we consider here - the dynamics at all times \cite{ferraro2005}. We also discussed the deviation of the steady state from the canonical Gibbs state as environmental coupling strength increases. Instead, the state of the system tends towards the \emph{mean force} Gibbs state \cite{trushechkin2022, grabert1988, fleming2011}. Finally, we saw how the quasiclassical approach can be applied to oscillator networks to effectively model quantum heat transport \cite{dhar2001, dhar2006, wang2007, wang2008, wang2009, segal2016, sevincli2019, li2021}. 

The exact agreement between the quasiclassical and open systems approaches arises here since the system of interest and coupling to the environment is linear \cite{schmid1982, grabert1984, grabert1988, eckern1990, hanggi2005}, but the correspondence in this regime provides a good platform for exploring non-linear setups. Particularly when the system of interest is strongly coupled to the environment, solving the full open quantum system can be computationally demanding \cite{strathearn2018, gribben2022, tanimura1989, jin2008, prior2010, chin2010, strasberg2016, nazir2018}, making techniques based on classical stochastic methods desirable. Indeed, in the field of molecular dynamics, such techniques have and are readily being explored \cite{ceriotti2009, dammak2009, ceriotti2010, ceriotti2011, ceriotti2016, lu2019}. The use of classical equations with `quantum' colored noise is also being employed to model magnetic materials, where quantum effects are becoming ever more relevant \cite{barker2019, spidy, berritta2024}.

\acknowledgments

We thank Simon Horsley, Stefano Scali, and Luis Correa for useful discussions on the subject of this article. The authors would also like to acknowledge the use of the University of Exeter High-Performance Computing (HPC) facility in carrying out this work. CRH is supported by a DTP grant from EPSRC (EP/T518049/1). JG is supported by a scholarship from CEMPS at the University of Exeter. FC and JA gratefully acknowledge funding from the Foundational Questions Institute Fund (FQXi-IAF19-01) and EPSRC (EP/R045577/1). JA thanks the Royal Society for support.

\section*{Author declarations}

\subsection*{Conflicts of Interest}

The authors have no conflicts to disclose.

\subsection*{Author Contributions}

\textbf{Charlie R. Hogg}: data curation; formal analysis (equal); software (equal); visualization; writing - original draft preparation (lead); writing - review and editing (equal).
\textbf{Jonas Glatthard}: formal analysis (equal); software (equal); writing - original draft preparation (supporting); writing - review and editing (equal). \textbf{Federico Cerisola}: formal analysis (supporting); software (supporting); supervision (supporting); visualization (supporting); writing - review and editing (equal).
\textbf{Janet Anders}: conceptualization; supervision (lead); writing - review and editing (equal).

\section*{Data availability statement}

\texttt{SpiDy.jl} is publicly available at \url{github.com/quantum-exeter/SpiDy.jl}. The code to produce Figs.~\ref{fig:noise}~-~\ref{fig:network} and associated data is available from CRH upon reasonable request.

\appendix

\section{Open quantum dynamics (OQu)} \label{sec:quantum_dynamics_appendix}

Here, we follow Ref.~\onlinecite{correa2023} to get the solution of the quantum dynamics. The Heisenberg equations of motion from the Hamiltonian $\op{H}_\sys + \op{H}_{\bath + \inter}$ defined in Eqs.~\eqref{eq:sys_hamiltonian}~-~\eqref{eq:int_hamiltonian} are
\begin{subequations} \label{eq:eqm}
    \begin{align}
        \dot{\op{x}} &= \frac{\op{p}}{m}, \\
        \dot{\op{p}} &= - m \, \wHOS^2 \, \op{x} - \int_0^\infty \dd\omega c_\omega \, \op{x}_\omega, \\
        \dot{\op{x}}_\omega &= \frac{\op{p}_\omega}{m_\omega}, \\
        \dot{\op{p}}_\omega &= - m_\omega \omega^2 \, \op{x}_\omega - c_\omega \, \op{x}.
    \end{align}
\end{subequations}
For compactness, we will define $\vect{\op{Z}}_\omega = (\op{x}_\omega, \op{p}_\omega)^\mathsf{T}$ and $\vect{\op{Z}} = (\op{x}, \op{p})^\mathsf{T}$, where vectors are denoted with an underline, and matrices with a double underline. The equations of motion for the environmental degrees of freedom can be expressed as
\begin{equation} \label{eq:vectorised_eom_bath}
    \dot{\vect{\op{Z}}}_\omega(t) + \mat{\wHO}_\omega \, \vect{\op{Z}}_\omega(t)  = \mat{C}_\omega \, \vect{\op{Z}}(t),
\end{equation}
with 
\begin{align}
	\mat{\wHO}_\omega &= \begin{pmatrix}
                                0 & - m_\omega^{-1} \\
                                m_\omega \, \omega^2&0
                             \end{pmatrix}, \\
	\mat{C}_\omega &= \begin{pmatrix}
                            0 & 0 \\
                            -c_\omega & 0
                          \end{pmatrix}.   
\end{align}
We take the Laplace transform of Eq.~\eqref{eq:vectorised_eom_bath} and obtain
\begin{equation}
    s \, \lap{{\vect{\op{Z}}}}_\omega(s) - \vect{\op{Z}}_\omega(0) + \mat{\wHO}_\omega \, \lap{{\vect{\op{Z}}}}_\omega(s) = \mat{C}_\omega \, \lap{{\vect{\op{Z}}}}(s),
\end{equation}
which can be expressed as
\begin{equation}
    \lap{{\vect{\op{Z}}}}_\omega(s) = \left( s \, \mat{\mathbbm{1}}_2 + \mat{\wHO}_\omega \right)^{-1} \left[ \mat{C}_\omega \, \lap{{\vect{\op{Z}}}}(s) + \vect{\op{Z}}_\omega(0) \right],
\end{equation}
where $\mat{\mathbbm{1}}_2$ is the $2 \times 2$ identity matrix. Now, taking the inverse Laplace transform, we get \cite{freitas2017}
\begin{equation}
    \vect{\op{Z}}_\omega(t) = \mat{G}_\omega(t)\, \vect{\op{Z}}_\omega(0) + \int_0^t \dd t' \, \mat{G}_\omega(t-t') \, \mat{C}_\omega \, \vect{\op{Z}}(t').
\end{equation}
Here, the elements of $G_\omega(t) $ are 
\begin{equation}
	\mat{G}_\omega(t) = \begin{pmatrix}
                                \cos{(\omega t)} & \frac{\sin{(\omega t)}}{m_\omega\,\omega} \\
                                -m_\omega \omega \sin{(\omega t)} & \cos{(\omega t)}
                            \end{pmatrix}.
\end{equation}
We can now express the equations of motion of the degrees of freedom of the system as \cite{weiss_book}
\begin{equation} \label{eq:vectorised_eom_sys}
    \dot{\vect{\op{Z}}}(t) + \mat{\wHO} \, \vect{\op{Z}}(t) - \int_0^t \dd t' \, \mat{\memker}(t-t') \, \vect{\op{Z}}(t') = \vect{\op{F}}(t),
\end{equation}
with
\begin{align}
	\mat{\Omega} &= \begin{pmatrix} 
                            0 & -\frac{1}{m} \\
                            m \, \wHOS^2 & 0
                        \end{pmatrix}, \\
	\mat{\memker}(t) &= \begin{pmatrix}
                                0 & 0 \\
                                \memker(t) & 0
                            \end{pmatrix},
\end{align}
where the noise term $\mat{\op{F}}(t) = \int_0^\infty \dd\omega \, \mat{C}_\omega \, \mat{G}_\omega(t) \, \vect{\op{Z}}_\omega(0)$. Note that the first line is Eq.~\eqref{eq:qu_langevin} from the main text. Again, taking the Laplace transform of \eqref{eq:vectorised_eom_sys} we obtain
\begin{equation}
    s \, \lap{{\vect{\op{Z}}}}(s) - \vect{\op{Z}}(0) + \mat{\wHO} \, \lap{{\vect{\op{Z}}}}(s) - \mat{\lap{\memker}}(s) \, \lap{{\vect{\op{Z}}}}(s) = \lap{\vect{\op{F}}}(s),
\end{equation}
which can be expressed as
\begin{equation}
    \lap{{\vect{\op{Z}}}}(s) = \left( s \, \mat{\mathbbm{1}}_2 + \mat{\wHO} - \mat{\lap{\memker}}(s) \right)^{-1} \left( \lap{\vect{\op{F}}}(s) + \vect{\op{Z}}(0) \right).
\end{equation}
Transforming this back into the time domain gives \cite{freitas2017}
\begin{equation} \label{eq:z}
    \vect{\op{Z}}(t) = \mat{G}(t) \, \vect{\op{Z}}(0) + \int_0^t \dd t' \, \mat{G}(t-t') \, \vect{\op{F}}(t'),
\end{equation}
with
\begin{equation}
    \mat{G}(t) = \begin{pmatrix}
                    g_1(t) & g_2(t) \\
                    g_3(t) & g_1(t)
                 \end{pmatrix}.
\end{equation}
The first moments of position and momentum are obtained by taking the expectation value over the elements of \eqref{eq:z} and, since $\expval{\op{F}(t)} = 0$, the first moments simply reduce to Eqs.~\eqref{eq:means} of the main text.

The second moments can be written as the covariance matrix
\begin{equation}
    \mat{\Sigma}(t) = \begin{pmatrix}
                        \sigma_{xx}(t) & \sigma_{xp}(t) \\
                        \sigma_{xp}(t) & \sigma_{pp}(t)
                      \end{pmatrix}, 
\end{equation}
which evolves as
\begin{multline} \label{eq:covariance_time}
    \mat{\Sigma}(t) = \mat{G}(t) \, \mat{\Sigma}(0) \, \mat{G}^\mathsf{T}(t) + \mat{G}(t) \, \Re{\expval{\vect{\op{Z}}(0) \, \vect{\op{B}}^\mathsf{T}(t)}} \\
    + \Re{\expval{\vect{\op{B}}(t) \, \vect{\op{Z}}^\mathsf{T}(0)}} \, \mat{G}^\mathsf{T}(t) + \Re{\expval{\vect{\op{B}}(t) \, \vect{\op{B}}^\mathsf{T}(t)}},
\end{multline}
with $\vect{\op{B}}(t)= \int_0^t \dd t' \, \mat{G}(t-t') \, \vect{\op{F}}(t')$. The second and third terms come from initial system-bath correlations and, therefore, vanish for our factorized initial state. The first term vanishes asymptotically but must be considered when studying the dynamics of the system. We now turn our attention to the last term, denoting it with a checkmark. Explicitly, this is given by
\begin{align}\label{eq:covariance_raw}
    \mat{\check\Sigma}(t) &= \Re{\expval{\vect{\op{B}}(t) \, \vect{\op{B}}^\mathsf{T}(t)}} \nonumber \\
    &= \int_0^t \dd t_1 \int_0^t \dd t_2 \int_0^\infty \dd\omega \int_0^\infty \dd\omega' \mat{G}(t-t_1) \mat{C}_\omega \, \mat{G}_\omega(t_1) \nonumber \\
    &\times \Re{\expval{\vect{\op{Z}}_\omega(0) \, \vect{\op{Z}}_{\omega'}^\mathsf{T}(0)}} \mat{G}_{\omega'}^\mathsf{T}(t_2) \, \mat{C}_{\omega'}^\mathsf{T} \, \mat{G}^\mathsf{T}(t-t_2).
\end{align}
Since the bath consists of oscillators in thermal equilibrium at temperature $T$, we have
\begin{multline}
    \Re{\expval{\vect{\op{Z}}_\omega(0) \, \vect{\op{Z}}_{\omega'}(0)^\mathsf{T}}} = \\
    \delta(\omega - \omega') \begin{pmatrix}
                                \frac{\hbar}{2 \, m_\omega \omega} \coth \left( \frac{\hbar \omega}{2 \kB T} \right) & 0 \\
                                0 & \frac{\hbar m_\omega \omega}{2} \coth \left( \frac{\hbar \omega}{2 \kB T} \right)
                             \end{pmatrix}, \label{eq:bath_corr} 
\end{multline}
where $\delta(\omega - \omega')$ is a delta function and the average $\expval{\op{A}} = \tr \left[ \op{A} \tau_B \right]$ here is taken over the initial thermal state $\op{\tau_B} = e^{- \beta \op{H_B}}/Z_B$ of the bath. Eq.~\eqref{eq:covariance_raw} thus simplifies to
\begin{equation} \label{eq:covariance_general}
    \mat{\check \Sigma}(t) = \frac{1}{2} \int_0^t \dd t_1 \int_0^t \dd t_2 \, \mat{G}(t-t_1) \begin{pmatrix}
                                                                                                0 & 0 \\
                                                                                                0 & \noiseker(t_1-t_2)
                                                                                             \end{pmatrix} \mat{G}(t-t_2)^\mathsf{T},
\end{equation}
with the noise autocorrelation function $\noiseker(\tau)$ given by Eq.~\eqref{eq:noise_autocorr_2}. Explicitly writing the elements of Eq.~\eqref{eq:covariance_general}, we have \cite{lampo_book}
\begin{subequations} \label{eq:covariance_general_elements}
    \begin{align}
        [\mat{\check\Sigma}(t)]_{11} &= \int_0^\infty \dd\omega \, \frac{\spd(\omega)}{\omega} \, \noise(\omega, T) \, \varsigma_{xx}(t, \omega), \\
        [\mat{\check\Sigma}(t)]_{12} &= \int_0^\infty \dd\omega \, \frac{\spd(\omega)}{\omega} \, \noise(\omega, T) \, \varsigma_{xp}(t, \omega), \\
        [\mat{\check\Sigma}(t)]_{22} &= \int_0^\infty \dd\omega \, \frac{\spd(\omega)}{\omega} \, \noise(\omega, T) \, \varsigma_{pp}(t, \omega),
    \end{align}
\end{subequations} 
with
\begin{subequations} \label{eq:varsigma}
    \begin{align}
        &\varsigma_{xx}(t, \omega) \nonumber \\
        &= \int_0^t \dd t_1 \int_0^t \dd t_2 g_2(t - t_1) \cos \left[ \omega(t_1 - t_2) \right] g_2(t - t_2) \nonumber \\
        &= \frac{1}{2} \int_0^t \dd u_1 g_2(u_1) e^{- i \, \omega \, u_1 } \int_0^t \dd u_2 g_2(u_2) e^{i \, \omega \, u_2} + \mathrm{c.c.}, \label{eq:varsigma_xx} \\
        &\varsigma_{xp}(t, \omega) \nonumber \\
        &= \int_0^t \dd t_1 \int_0^t \dd t_2 g_2(t - t_1) \cos \left[ \omega(t_1 - t_2) \right] g_1(t - t_2) \nonumber \\
        &= \frac{1}{2} \int_0^t \dd u_1 g_2(u_1) e^{- i \, \omega \, u_1 } \int_0^t \dd u_2 g_1(u_2) e^{i \, \omega \, u_2} + \mathrm{c.c.}, \label{eq:varsigma_xp} \\
        &\varsigma_{pp}(t, \omega) \nonumber \\
        &= \int_0^t \dd t_1 \int_0^t \dd t_2 g_1(t - t_1) \cos \left[ \omega(t_1 - t_2) \right] g_1(t - t_2) \nonumber \\
        &= \frac{1}{2} \int_0^t \dd u_1 g_1(u_1) e^{- i \, \omega \, u_1 } \int_0^t \dd u_2 g_1(u_2) e^{i \, \omega \, u_2} + \mathrm{c.c.}, \label{eq:varsigma_pp}
    \end{align}
\end{subequations}
where we have used Euler's formula, $\cos x = \frac{1}{2} (e^{i \, x} + e^{- i \, x})$, and introduced $u_j = t - t_j$. Here, $\mathrm{`c.c.'}$ denotes the complex conjugate. This gives us the first terms in Eqs.~\eqref{eq:covs} in the main text with quantum noise $\noise(\omega, T) = \noiseQu(\omega, T)$. The second term, involving the initial conditions, comes from the first term in Eq.~\eqref{eq:covariance_time}, i.e. $\mat{G}(t) \, \mat{\Sigma}(0) \, \mat{G}^\mathsf{T}(t)$.

\section{Open classical dynamics (OCl)} \label{sec:classical_dynamics_appendix}

We look at the same system but classically. Instead of operators, we now have phase-space variables and functions. So, the Hamiltonians \eqref{eq:sys_hamiltonian} and \eqref{eq:int_hamiltonian} become
\begin{align}
    H_\sys &= \frac{1}{2} m \wHO^2 \, x^2 + \frac{1}{2m} \, p^2, \label{eq:sys_Hamiltonian_cl} \\
    H_{\bath + \inter} &= \int_0^\infty \dd\omega \frac{1}{2 m_\omega} \left[ p_\omega^2 + \left( m_\omega \omega \, x_\omega + \frac{c_\omega}{\omega} x \right)^2 \right],
\end{align}
respectively. We consider the same spectral density $\spd(\omega)$ as above. The Liouville equations of motion are
\begin{align}
    \dot{x}(t) &= \frac{p(t)}{m}, \\
    \dot{p}(t) &= - m \, \wHOS^2 \, x(t) - \int_0^\infty \dd\omega \, c_\omega x_\omega(t), \\
    \dot{x}_\omega(t) &= \frac{p_\omega(t)}{m_\omega}, \\
    \dot{p}_\omega(t) &= - m_\omega \omega^2 \, x_\omega(t) - c_\omega \, x(t).
\end{align}
As the equations of motion are the same as in the quantum case, we can follow the derivation above. We arrive at the classical Langevin equation
\begin{equation}
    m \ddot{x}(t) + m \wHOS^2 \, x(t) - \int_0^t \dd t' \, \memker(t - t') \, x(t') =  F(t),
\end{equation}
where $\memker(t - t')$ is the same as above, and $F(t)$ takes the form
\begin{equation}
    F(t) = - \int_0^\infty \dd\omega \, c_\omega \left(  x_\omega(0) \cos{(\omega t)} + \frac{\ p_\omega(0)}{m_\omega \omega} \sin{(\omega t)} \right).
\end{equation}
We end up with the same equation as in the classical case but for variables instead of operators. As the equations of motion are the same as in the quantum case, we can follow the derivation above. The only difference is with Eq.~\eqref{eq:bath_corr}, which in the classical case reads
\begin{equation}
    \Re{\expval{Z_\omega(0) \, Z_{\omega'}(0)^\mathsf{T}}} = \delta(\omega - \omega') \begin{pmatrix}
                                                                                            \frac{\kBT}{m_\omega \, \omega^2} & 0 \\
                                                                                            0 & m_\omega \, \kBT
                                                                                        \end{pmatrix}.
\end{equation}
Going through the analogous steps as in the quantum case, one recovers Eqs.~\eqref{eq:means}~-~\eqref{eq:covs} of the main text, but with classical noise $\noise(\omega, T) = \noiseCl(\omega, T)$.

\section{Stochastic dynamics (StoQu and StoCl)} \label{sec:stochastic_dynamics_appendix}

To solve for both the stochastic classical (StoCl) and quantum (StoQu) dynamics, we begin by applying a Laplace transform to Eqs.~\eqref{eq:sto_eqm}. This gives
\begin{subequations} \label{eq:sto_eqm_lt}
    \begin{align}
        s \lap{x}(s) - x(0) &= \frac{\lap{p}(s)}{m}, \label{eq:sto_eqm_1_lt} \\
        s \lap{p}(s) - p(0) &= - m \, \wHOS^2 \lap{x}(s) + \lap{F}(s) + \lap{\memker}(s) \, \lap{x}(s). \label{eq:sto_eqm_2_lt}
    \end{align}
\end{subequations}
After rearranging and substitution, we get the following expression for $\lap{x}(s)$
\begin{equation} \label{eq:sto_x_s}
    \lap{x}(s) = \lap{g}_2(s) \lap{F}(s) + \lap{g}_1(s) \, x(0) + \lap{g}_2(s) \, p(0).
\end{equation}
Applying the inverse Laplace transform, we have
\begin{equation} \label{eq:sto_x_t}
    x(t) = \invlap \left[ \lap{g}_2(s) \lap{F}(s) \right] + g_1(t) \, x(0) + g_2(t) \, p(0).
\end{equation}
We now take the time derivative to obtain the momentum
\begin{equation} \label{eq:sto_p_t}
    p(t) = m \, \partial_t \invlap \left[ \lap{g}_2(s) \lap{F}(s) \right] + g_3(t) \, x(0) + g_1(t) \, p(0),
\end{equation}
where we have used the fact that $g_3(t) = m \, \partial_t g_1(t)$, and $g_1 (t) = m \, \partial_t g_2(t)$. Utilizing the convolution theorem, we can reexpress the inverse Laplace transforms in Eqs.~\eqref{eq:sto_x_t}~and~\eqref{eq:sto_p_t} as
\begin{align}
    x(t) &= \int_0^t \dd t' \, g_2(t - t') F(t') + g_1(t) \, x(0) + g_2(t) \, p(0), \label{eq:sto_x_t2} \\
    p(t) &= m \, \int_0^t \dd t' \, g_1(t - t') F(t') + g_3(t) \, x(0) + g_1(t) \, p(0), \label{eq:sto_p_t2}
\end{align}
where we have used the fact \cite{riley_book} that $\partial_x (f \ast g) = \partial_x f \ast g = f \ast \partial_x g$. When we take a stochastic average of $x$ and $p$, the first terms of Eqs.~\eqref{eq:sto_x_t2}~and~\eqref{eq:sto_p_t2} will disappear since $\expval{F(t)}_\sto = 0$ over all realizations at any given time $t$. Taking a further average over the initial conditions then yields Eqs.~\eqref{eq:means} of the main text with $\mu_q(0) = \expval{q_0}_0, \, q \in \{ x, p \}$.

To find the second moments, we first take the two-time average of the products, utilizing the fact that the time average over any \emph{single} $F(t)$ term will be zero. This gives
\begin{widetext}
    \begin{align}
        \begin{split}
            \llangle x(t_1) x(t_2) \rrangle_0 &= \int_0^{t_1} \dd t_1' \int_0^{t_2} \dd t_2' \, g_2(t_1 - t_1') g_2(t_2 - t_2') \expval{F(t_1') F(t_2')}_\sto \\ 
            & \qquad + g_1(t_1) g_1(t_2) \expval{x(0)^2}_0 + \left( g_1(t_1) g_2(t_2) + g_1(t_2) g_2(t_1) \right) \expval{x(0) p(0)}_0 + g_2(t_1) g_2(t_2) \expval{p(0)^2}_0, \label{eq:sto_mean_xt1_xt2}
        \end{split} \\
        \begin{split}
            \llangle x(t_1) p(t_2) \rrangle_0 &= m \int_0^{t_1} \dd t_1' \int_0^{t_2} \dd t_2' \, g_2(t_1 - t_1') g_1(t_2 - t_2') \expval{F(t_1') F(t_2')}_\sto \\
            & \qquad + g_1(t_1) g_3(t_2) \expval{x(0)^2}_0 + \left( g_1(t_1) g_1(t_2) + g_2(t_1) g_3(t_2) \right) \expval{x(0) p(0)}_0 + g_1(t_2) g_2(t_1) \expval{p(0)^2}_0, \label{eq:sto_mean_xt1_pt2}
        \end{split} \\
        \begin{split}
            \llangle p(t_1) p(t_2) \rrangle_0 &= m^2 \int_0^{t_1} \dd t_1' \int_0^{t_2} \dd t_2' \, g_1(t_1 - t_1') g_1(t_2 - t_2') \expval{F(t_1') F(t_2')}_\sto \\
            & \qquad + g_3(t_1) g_3(t_2) \expval{x(0)^2}_0 + \left( g_1(t_1) g_3(t_2) + g_1(t_2) g_3(t_1) \right) \expval{x(0) p(0)}_0 + g_1(t_1) g_1(t_2) \expval{p(0)^2}_0. \label{eq:sto_mean_pt1_pt2}
        \end{split}
    \end{align}
\end{widetext}
Rearranging \eqref{eq:sto_psd}, we have
\begin{equation} \label{eq:sto_corr}
    \expval{F(t) \, F(t - \tau)}_\sto = \int_0^\infty \dd\omega \frac{\spd(\omega)}{\omega} \noise(\omega, T) \cos(\omega\tau).
\end{equation}
where $\noise(\omega, T) = \noiseQu(\omega, T)$ (StoQu) or $\noise(\omega, T) = \noiseCl(\omega, T)$ (StoCl), which can then be substituted into Eqs.~\eqref{eq:sto_mean_xt1_xt2}~-~\eqref{eq:sto_mean_pt1_pt2}. Setting $t_1 = t_2 = t$ gives exactly the first lines of Eqs.~\eqref{eq:varsigma}. Following the same steps as in Appendix~\ref{sec:quantum_dynamics_appendix} recovers Eqs.~\eqref{eq:covs}~and~\eqref{eq:I} of the main text, where we have used the fact that $\sigma_{xp}(t) = \sigma_{px}(t)$ since, here, the position and momentum are classical variables that commute at all times. 

\section{Steady state} \label{sec:steady_state_appendix}

Here, we take the long-time limit of the covariance matrix. We find that the first term in Eq.~\eqref{eq:covariance_time} vanishes, as $\mat{G}(t)$ asymptotically vanishes. The remaining terms,  Eqs.~\eqref{eq:covariance_general_elements}, can be studied through $\varsigma_{xx}(t, \omega), \varsigma_{xp}(t, \omega)$ and $\varsigma_{pp}(t, \omega)$ in Eqs.~\eqref{eq:varsigma}.
In the long-time limit, we have
\begin{equation} \label{eq:varsigma_xx_ss}
    \lim_{t \to \infty} \, \varsigma_{xx}(t, \omega) = \lap{g}_2(- i \, \omega) \, \lap{g}_2(i \, \omega). 
\end{equation}
Furthermore, using $\lap{g}_1(s) = m \, s \, \lap{g}_2(s)$, we find
\begin{align} \label{eq:varsigma_xp_ss}
    \lim_{t \to \infty} \, \varsigma_{xp}(t, \omega) = \frac{m}{2} \left[ i \, \omega \, \vert \lap{g}_2(i \, \omega) \vert^2 - i \, \omega \, \vert \lap{g}_2(i \, \omega) \vert^2 \right] = 0.
\end{align}
And finally
\begin{align} \label{eq:varsigma_pp_ss}
    \lim_{t \to \infty} \, \varsigma_{pp}(t, \omega) = m^2 \, \omega^2 \, \vert\lap{g}_2(i \, \omega) \vert^2.
\end{align}
Substitution of \eqref{eq:varsigma_xx_ss}~-~\eqref{eq:varsigma_pp_ss} into
Eqs.~\eqref{eq:covariance_general_elements} gives exactly Eqs.~\eqref{eq:covs_ss} of the main text with $\noise(\omega, T) = \noiseQu(\omega, T)$ for the OQu and StoQu approaches, and $\noise(\omega, T) = \noiseCl(\omega, T)$ for the OCl and StoCl approaches. Again, note the only difference between the quantum and classical models is the form of the noise term $\noise(\omega, T)$.

One can also directly obtain the steady-state covariances \eqref{eq:covs_ss} by considering the power spectral density of the environmental force \cite{breuer_book}. Note that whilst we consider the classical equations of motion below (for the StoQu dynamics), this is also applicable to the StoCl, OQu and OCl models. As an example, we will evaluate $\sigma_{xx}(\infty)$. First, we take the Fourier transform of Eqs.~\eqref{eq:sto_eqm}, obtaining
\begin{subequations} \label{eq:sto_eqm_ft}
    \begin{align}
        i \omega \, \fourier{x}(\omega) &= \frac{\fourier{p}(\omega)}{m} \label{eq:sto_eqm1_ft} \\
        i \omega \, \fourier{p}(\omega) &= - m \, \wHOS^2 \fourier{x}(\omega) + \fourier{F}(\omega) + \fourier{\memker}(\omega) \, \fourier{x}(\omega), \label{eq:sto_eqm2_ft}
    \end{align}
\end{subequations}
where $\fourier{x}(\omega)$ represents the Fourier transform of $x(t)$ (and similarly for the other functions). We are then simply left with a linear set of equations that can be readily solved to obtain $\fourier{x}(\omega)$ as a function of the stochastic field $\fourier{F}(\omega)$
\begin{equation} \label{eq:sto_x_omega}
    \fourier{x}(\omega) = \frac{\fourier{F}(\omega)}{m \left( \wHOS^2 - \omega^2 \right) - \fourier{\memker}(\omega)}.
\end{equation}
This can be used to readily relate the power spectral density of $x(t)$, $\psd_x(\omega) = \abs{\fourier{x}(\omega)}^2$, with that of $F$ via
\begin{equation}
    \psd_x(\omega) = \frac{\psd_F(\omega)}{\abs{m \left( \wHOS^2 - \omega^2 \right) - \fourier{\memker}(\omega)}^2}.
\end{equation}
As discussed, the power spectral density of $F$ is given by \eqref{eq:sto_psd}. Therefore, the position auto-correlation function is
\begin{align}
    \expval{x(t) \, x(t - \tau)} &= \frac{1}{2 \pi} \int_{-\infty}^{\infty} \dd\omega \, \psd_x(\omega) e^{-i \omega \tau} \nonumber \\
    &= \frac{1}{2} \int_{-\infty}^{\infty} \dd\omega \, \frac{\spd(\omega) \noiseQu(\omega, T) e^{-i \omega \tau}}{\omega \, \abs{m \left( \wHOS^2 - \omega^2 \right) - \fourier{\memker}(\omega)}^2}.
\end{align}
In particular, since $\expval{x(t)} = 0$ in the steady state, the variance is given by
\begin{align} \label{eq:sto_cov_xx_ss}
    \sigma_{xx}(\infty) &= \int_{0}^{\infty} \dd\omega \, \frac{\spd(\omega) \noiseQu(\omega, T)}{\omega \, \abs{m \left( \wHOS^2 - \omega^2 \right) - \fourier{\memker}(\omega)}^2} \nonumber \\
    &= \int_0^{\infty} \dd\omega \frac{\spd(\omega) \noiseQu(\omega, T)}{\omega \, \abs{m \left( \wHOS^2 + (i \omega)^2 \right) - \lap{\memker}(i \omega)}^2},
\end{align}
which is readily seen to be equal to Eq.~\eqref{eq:cov_xx_ss} derived for the full quantum treatment of harmonic oscillator and bath.

\section{Exact expressions} \label{sec:exact_appendix}

Here, we give exact expressions for the steady-state covariances for the case of a Lorentzian spectral density, Eq.~\eqref{eq:spectral_density_lor}. We note that the integrands in Eqs.~\eqref{eq:covs_ss} are symmetric. The integral is, therefore, the same as half the integral from $-\infty$ to $\infty$ of the same integrand. Using the identity $\hbar \coth \left( \hbar \omega / 2 \kBT \right) = 2 \sum_{n=1}^\infty \frac{2 \kBT \omega}{\nu_n^2 + \omega^2} + 2 \frac{\kBT}{\omega}$, where $\nu_n := 2 \pi \kBT n / \hbar$ are the Matsubara frequencies, we can rewrite the integral in Eq.~\eqref{eq:cov_xx_ss} as \cite{grabert1988, hanggi2005}
\begin{align}
    \sigma_{xx} &= \frac{\GammaLor \lambda^2 T}{\pi} \Bigg( \int_{-\infty}^\infty \dd\omega \, \sum_{n=1}^\infty \frac{2\omega^2}{h_q^{(n)}(\omega ) \, h_q^{(n)}(-\omega )} \nonumber \\ 
    &+ \int_{-\infty}^\infty \dd\omega \, \frac{1}{h_c(\omega )\,h_c(-\omega )} \Bigg),
\end{align}
where $h_q^{(n)}(\omega) = h_c(\omega) (\nu_n +  i \omega)$ and $h_c(\omega)= - m \omega^4 + i m \GammaLor \omega^3 + m \wHO^2 \omega^2  + m \wLor^2 \omega^2 + \lambda^2 \omega^2 / \wLor^2  - i m \GammaLor \wHO^2 \omega - i \GammaLor \lambda^2 \omega / \wLor^2 - m \wHO^2 \wLor^2$. In the classical case, only the second integral appears. Such integrals can be evaluated using the formula \cite{gradshteyn_book}
\begin{equation}
    \int_{-\infty}^\infty \dd x \, \frac{g_n(x)}{h_n(x) h_n(-x)} = (-1)^{(n+1)} \frac{i \pi}{a_0} \frac{\det \mathsf{M}_n}{\det \mathsf{\Delta}_n},
\end{equation}
where $g_n(x) = b_0 x^{2n-2} + b_1 x^{2n-4} +\dots+b_{n-1}$ and $h_n(x) = a_0 x^n + a_1 x^{n-1} +\dots+a_n$ and the matrices $\mathsf{M}_n$ and $\mathsf{\Delta}_n$ are defined as
\begin{equation}
    \mathsf{M}_n = \begin{bmatrix} 
                       b_0 & b_1 & \dots & b_{n-1} \\
                       a_0 & a_2 & \dots & 0 \\
                       0 & a_1 & \dots & 0 \\
                       \vdots & \vdots & \ddots & \vdots \\
                       0 & 0 &\hdots & a_n 
                   \end{bmatrix},
\end{equation}
and
\begin{equation}
    \mathsf{\Delta}_n = \begin{bmatrix} 
                            a_1 & a_3 & \dots & 0 \\
                            a_0 & a_2 & \dots & 0 \\
                            0 & a_1 & \dots & 0 \\
                            \vdots & \vdots & \ddots & \vdots \\
                            0 & 0 &\hdots & a_n 
                        \end{bmatrix}.
\end{equation}
To avoid confusion, we note that the last row of both matrices ends with $\dots, a_{n-4}, a_{n-2}, a_n$. For the formula to be valid, all the roots of $h_n(x)$ must lie in the upper half of the complex plane, which is the case here. The analogous procedure can be applied to $\sigma_{pp}$. Lastly, $\sigma_{xp}$ vanishes due to anti-symmetry. We get the following expressions (for the sake of brevity in nondimensional units, i.e. $\hbar = \kB = m = \wHO = 1$)
\begin{subequations}\label{eq:covariances_coefficients_cl_simplified}
    \begin{align}
        &\sigma_{xx}(\infty) = T \nonumber \\
        & + \sum_{n=1}^\infty \frac{2 T \wLor^2 (\nu_n (\GammaLor + \nu_n) + \wLor^2)}{\lambda^2 \nu_n (\GammaLor + \nu_n) + \nu_n(\GammaLor + \nu_n)(1 + \nu_n^2) \wLor^2 + (1 + \nu_n^2) \wLor^4}, \\
        &\sigma_{xp}(\infty) = 0, \\
        &\sigma_{pp}(\infty) = T \nonumber \\
        & + \sum_{n=1}^\infty \frac{2 T (\lambda^2 \nu_n (\GammaLor + \nu_n)+ \nu_n(\GammaLor + \nu_n) \wLor^2 + \wLor^4)}{\lambda^2 \nu_n (\GammaLor + \nu_n)+ \nu_n(\GammaLor + \nu_n)(1 + \nu_n^2) \wLor^2 + (1 + \nu_n^2) \wLor^4}.
    \end{align}
\end{subequations}
Using the same method for the classical case, reintroducing units yields 
\begin{subequations}\label{eq:covariances_coefficients_cl_simplified_2}
    \begin{align}
        \sigma_{xx}(\infty) &= \frac{\kBT}{m \wHO^2}, \\
        \sigma_{xp}(\infty) &= 0, \\
        \sigma_{pp}(\infty) &= m \kBT,
    \end{align}
\end{subequations}
which corresponds to the thermal state. 

Without the counter-term, both for the quantum and classical cases, $\wHOS^2$ in Eq.~\eqref{eq:qu_langevin} would read $\wHO^2$. The above analysis would, therefore, be valid as long as this shift is taken into account. For the Lorentzian spectral density, this results in the shifted frequency $\sqrt{\wHO^2 -  \lambda^2/m \wLor^2}$. The steady state of the dissipative oscillator then is still a thermal state but with this shifted frequency. This is in line with the classical mean force corrections, which only include the reorganization energy \cite{cresser2021, timofeev2022, correa2023, cerisola2024}.

\section{Network} \label{sec:network_appendix}

Following analogous steps as for a single quantum oscillator, the equations of motion for the quantum network, both in the open systems and stochastic cases, can be cast into the form
\begin{subequations} \label{eq:neom}
\begin{align} 
    \dot{\vect{\op{X}}}(t) &= \mat{M}^{-1} \vect{\op{P}}(t), \\
    \dot{\vect{\op{P}}}(t) &= - \mat{\bar{V}} \, \vect{\op{X}}(t) + \int_0^t \dd t' \, \mat{\memker}(t - t') \, \vect{\op{X}}(t') + \vect{\op{F}}(t). \label{eq:neomb}
\end{align}
\end{subequations}
The noise term $\vect{\op{F}}(t)$ is a column vector, with entries microscopically given by
\begin{multline}
    \op{F}_\gamma(t) = - \sum_\alpha \int_0^\infty \dd\omega \, C_{\omega,\alpha \gamma} \\
    \times \left( \op{x}_{\omega,\alpha}(0) \cos(\omega t) + \frac{\op{p}_{\omega,\alpha}(0)}{m_{\omega,\alpha} \omega} \sin (\omega t) \right),
\end{multline}
and the dissipation kernel matrix is given by $\mat{\memker}(\tau) = 2 \int_0^\infty \dd\omega \sum_\alpha \mat{\spd}_\alpha(\omega) \sin (\omega \tau)$. The components of the spectral density are given by
\begin{equation}
    \spd_\alpha(\omega)_{\gamma, \delta} = \frac{1}{2 m_{\omega, \alpha} \omega} C_{\omega, \alpha \gamma} \, C_{\omega, \alpha \delta},
\end{equation}
and result in the power spectrum
\begin{equation}
    \mat{\psd}^\qu_F(\omega) = \pi \sum_\alpha \frac{\mat{\spd}_\alpha(\omega)}{\omega} \noisei^\qu(\omega).
\end{equation}
The baths can have different individual power spectra $\mat{\spd}_\alpha(\omega)$ and can be initialised at different temperatures $T_\alpha$. Similarly we can write $\mat{\memker}(\tau) = \sum_\alpha \mat{\memker}_\alpha(\tau)$ and $ \mat{\bar{V}} = \mat{V} + \sum_\alpha \mat{\Delta V}_\alpha$, where $\mat{\memker}_\alpha$ and $\mat{\Delta V}_\alpha$ can be obtained by $\mat{\spd}_\alpha(\omega)$ as in the single bath case.  

Denoting $\vect{\op{Z}} = (\vect{\op{X}}, \vect{\op{P}})^\mathsf{T}$, the equations of motions can be solved using the Laplace transform by
\begin{equation}
    \vect{\op{Z}}(t) = \mat{G}(t) \, \vect{\op{Z}}(0) + \int_0^t \dd t' \, \mat{G}(t - t') \, \vect{\op{F}}(t'),  
\end{equation}
where $\vect{\op{F}}(t')$ is a column vector with odd entries being equal to zero and the $2 \gamma$-th entries  being $\op{F}_\gamma(t)$ and $\mat{G}(s)$ is the inverse Laplace transform of
\begin{equation}
    \lap{\mat{G}}(s)= \left( s \, \mat{\mathbbm{1}}_{2N} + \mat{\Omega} - \lap{\mat{K}}(s) \right)^{-1},  
\end{equation}
with
\begin{align}
    \mat{\Omega} &= \begin{pmatrix}
                        0 & -\mat{M}^{-1} \\
                        \mat{\bar{V}} & 0
                    \end{pmatrix}, \\
    \lap{\mat{K}}(s) &= \begin{pmatrix}
                            0 & 0 \\ 
                            \lap{\mat{\memker}}(s) & 0
                        \end{pmatrix}.
\end{align}
Due to linearity, the averages of the correlation functions can be calculated by averages over the initial state and noise term. The symmetrized correlation functions of the open quantum system dynamics, therefore, match the stochastic one with appropriate correlation functions of the noise, i.e. the quasiclassical quantum noise as in Eq.~\eqref{eq:qu_noise}.

\section{Heat currents} \label{sec:heat_current_appendix}

The heat currents can be calculated by the average change of energy due to the interaction with the baths \cite{freitas2017}. We first calculate the change of energy and then identify the contributions of each bath. By only making use of system operators and the system equation of motion, the derivation, therefore, equally applies to the open systems, stochastic, quantum and classical cases. The change of energy can be written as
\begin{align} \label{eq:energy_change}
    \expval{\dot{\op{H}}_{\sys}(t)} &= \frac{1}{2} \expval{ \frac{d}{dt} \left( \vect{\op{P}}^\mathsf{T}(t) \mat{M}^{-1} \vect{\op{P}}(t) + \vect{\op{X}}^\mathsf{T}(t) \mat{\bar{V}} \, \vect{\op{X}}(t) \right) } \nonumber \\
    &= \frac{1}{2} \expval{ \frac{d}{dt} \tr \left( \vect{\op{P}}(t) \left( \mat{M}^{-1} \vect{\op{P}}(t) \right)^\mathsf{T} + \vect{\op{X}}(t) \left( \mat{\bar{V}} \,\vect{\op{X}}(t) \right)^\mathsf{T} \right) } \nonumber \\
    &= \frac{1}{2} \expval{ \frac{d}{dt} \tr \left( \vect{\op{P}}(t) \vect{\op{P}}^\mathsf{T}(t) \mat{M}^{-1} + \vect{\op{X}}(t) \vect{\op{X}}^\mathsf{T}(t) \mat{\bar{V}} \right) },
\end{align}
where $\tr()$ is the trace taken over matrices containing the components of the different oscillators. Therefore, we only have to consider the derivatives of the positions and momenta to get the change in energy.

The heat currents are calculated by the change due to the interaction with a specific bath $\alpha$. As seen in Eqs.~\eqref{eq:neom}, $\dot{\vect{\op{X}}}(t)$ does not include terms involving the baths, while each bath contributes to $\dot{\vect{\op{P}}}(t)$ as $\vect{\op{F}}_\alpha(t) + \int_0^t \dd t' \, \mat{\memker}_\alpha(t - t') \vect{\op{X}}(t')$. In the following, we assume that the baths couple to separate oscillators, i.e. 
\begin{subequations}
    \begin{align}
        \vect{\op{F}}_\alpha(t) &= \mat{\Pi}_\alpha \, \vect{\op{F}}(t),  \\
        \mat{\memker}_\alpha(t) &= \mat{\Pi}_\alpha \, \mat{\memker}(t), 
    \end{align}
\end{subequations}
where $\mat{\Pi}_\alpha$ is the projector to the oscillators which the bath $\alpha$ couples to. We can therefore write the contribution to $\dot{\vect{\op{P}}}(t)$ by the specific bath $\alpha$ by applying the projector $\mat{\Pi}_\alpha$ to Eq.~\eqref{eq:neomb} as
\begin{equation}
    \vect{\op{F}}_\alpha(t) + \int_0^t \dd t' \, \mat{\memker}_\alpha(t - t') \vect{\op{X}}(t') = \mat{\Pi}_\alpha \left( \dot{\vect{\op{P}}}(t) + \mat{\bar{V}} \, \vect{\op{X}}(t) \right).
\end{equation}

Considering only the terms due to a specific bath in the last line of Eq.~\eqref{eq:energy_change}, we get the heat current by using the product rule, i.e.
\begin{align} \label{eq:heat_current}
    \dot Q_\alpha(t) =& \frac{1}{2} \bigg\langle \tr \left( \mat{\Pi}_\alpha \left( \dot{\vect{\op{P}}}(t) + \mat{\bar{V}} \vect{\op{X}}(t) \right) \vect{\op{P}}^\mathsf{T}(t) \mat{M}^{-1} \right. \nonumber \\
    &+ \left. \vect{\op{P}}(t) \left[ \mat{\Pi}_\alpha \left( \dot{\vect{\op{P}}}(t) + \mat{\bar{V}} \vect{\op{X}}(t) \right) \right]^\mathsf{T} \mat{M}^{-1} \right) \bigg\rangle \nonumber \\
    =& \frac{1}{2} \tr \left[ \mat{\Pi}_\alpha \, \dot{\mat{C}}_{pp}(t) \mat{M}^{-1} \right] + \tr \left[ \mat{\Pi}_\alpha \, \mat{\bar{V}} \, \mat{C}_{xp}(t) \mat{M}^{-1} \right].
\end{align}
Here, we denoted $\mat{C}_{pp}(t) = \langle \vect{\op P}(t) \vect{\op P}(t)^\mathsf{T} \rangle$ and $\mat{C}_{xp}(t) = \frac{1}{2}\langle \{\vect{\op X}(t),\vect{\op P}(t)^\mathsf{T}\} \rangle$. Again, the trace is taken over matrices containing the components of the different oscillators. In the long-time limit, only the second term gives a contribution as the covariance matrices approach their stationary values.

\medskip

For the stochastic treatment, since we now have variables instead of operators, we can also express \eqref{eq:heat_current} as
\begin{align} \label{eq:heat_current_sto_2}
    \dot Q_\alpha(t) &= \frac{1}{2} \bigg \langle \tr \bigg( \left( \vect{F}_\alpha(t) + \vect{u}_\alpha(t) \right) \vect{P}^\mathsf{T}(t) \mat{M}^{-1} \nonumber \\
    & \qquad \qquad \qquad + \vect{P}(t) \left( \vect{F}_\alpha(t) + \vect{u}_\alpha(t) \right)^\mathsf{T} \mat{M}^{-1} \bigg) \bigg \rangle \nonumber \\
    &= \frac{1}{2} \bigg\langle \left( \vect{F}_\alpha(t) + \vect{u}_\alpha(t) \right)^\mathsf{T} \mat{M}^{-1} \vect{P}(t) \nonumber \\
    & \qquad \qquad \qquad + \vect{P}^\mathsf{T}(t) \mat{M}^{-1} \left( \vect{F}_\alpha(t) + \vect{u}_\alpha(t) \right) \bigg\rangle,
\end{align}
where $\vect{u}_\alpha(t) = \vect{F}_\alpha(t) + \int_0^t \dd t' \, \mat{\memker}_\alpha(t - t') \vect{x}(t')$. Since $\vect{P}(t) = \mat{M} \, \vect{\dot{X}}(t)$, we can write
\begin{align} \label{eq:heat_current_sto_3}
    \dot Q_\alpha(t) &= \frac{1}{2} \big\langle \left( \vect{F}_\alpha(t) + \vect{u}_\alpha(t) \right)^\mathsf{T} \vect{\dot{X}}(t) + \vect{\dot{X}}^\mathsf{T}(t) \left( \vect{F}_\alpha(t) + \vect{u}_\alpha(t) \right) \big\rangle \nonumber \\
    &= \left\langle \left(\vect{F}_\alpha(t) + \vect{u}_\alpha(t) \right) \cdot \vect{\dot{X}}(t) \right\rangle,
\end{align}
which is Eq.~\eqref{eq:sto_currents} of the main text.

\bibliography{refs}

\end{document}